\documentstyle[eqsecnum,aps,epsfig]{revtex}
\begin{document}
\draft
\preprint{hep-ph/0010322}
\title{
MUM: flexible precise Monte Carlo algorithm for muon propagation
through thick layers of matter 
}
\author{Igor A. Sokalski, Edgar V. Bugaev, Sergey I. Klimushin}
\address{
Institute for Nuclear Research, Russian Academy of Science, 
60th October Anniversary prospect 7a, Moscow 117312, Russia
}
\date{\today}
\maketitle
\begin{abstract}
We present a new Monte Carlo muon propagation algorithm MUM (MUons+Medium) 
which possesses some advantages over analogous algorithms presently in use. The
most important features of algorithm are described. Results on the test for 
accuracy of treatment the muon energy loss with MUM are presented and analyzed.
It is evaluated to be of 2$\times$10$^{-3}$ or better, depending upon 
simulation parameters. Contributions of different simplifications which are 
applied at Monte Carlo muon transportation to the resulting error are 
considered and ranked. It is shown that when simulating muon propagation 
through medium it is quite enough to account only for fluctuations in radiative
energy loss with fraction of energy lost being as large as 0.05$\div$0.1. 
Selected results obtained with MUM are given and compared with ones from other 
algorithms.  
\end{abstract}

\widetext

\section{INTRODUCTION}
\label{sec:int1}

Muon propagation through thick layers of matter was in the scope of interest 
for a long time since the first underground experiments with natural muon and
neutrino fluxes had started. Development of ``underground'' technique has led
to creation of number of underground, underwater and under-ice detectors by 
which a wide spectrum of problems is presently under investigation. Accurate 
calculation of muon transport plays an important role for such experiments 
because {\it (a)} neutrinos are detected by muons which are born in $\nu N$
interactions and propagate a distance in medium from the point of interaction 
to a detector;  {\it (b)} muons which are produced in atmospheric showers
generated by cosmic rays represent the principal background for neutrino signal
and therefore their flux at large depths should be well known; {\it (c)} 
atmospheric muons deep under sea or earth surface are the only intensive and 
more or less known natural calibration source which allows to confirm 
correctness of the detector model by comparison experimental and expected 
detector response; {\it (d)} flux of atmospheric muons itself carries the 
physical information which is of interest. 

Along with analytical and semi-analytical methods 
(Refs.~\cite{ZM62,K67,H69,KKO72,MN76,GZM76,MKK81,BNS85IAN,B86,AB86PR,BNNSC92,N94,naumov1})
one widely uses Monte Carlo (MC) technique 
(Refs.~\cite{HPW63,OWY68,BP71,VTF74,TKAOM84,lohman,CBAPBP85,BCdEPMPV91,lipari,M92,music1,lagutin1})
which directly accounts for the stochastic nature of muon energy losses to 
simulate the muon propagation through matter. There are several MC muon 
transportation algorithms currently in use (see, e.g. Ref.~\cite{rhode1} for 
detailed analysis of their advantages and disadvantages), but essential 
theoretical and experimental progress of last years makes to create new ones. 
Here we present a MC muon propagation code MUM (MUons+Medium) written in 
FORTRAN which possesses some advantages in accuracy and flexibility over 
analogous simulation algorithms (although it does not contain some important 
features in its current version, e.g. it does not give the 3D information about
angular and lateral deviations of muons). The algorithm has been developed for 
the Baikal deep underwater neutrino experiment (Ref.~\cite{NT1}) but we believe
it to be useful also for other experiments with natural fluxes of high energy 
muons and neutrinos. When working on MUM we aimed at creation of an algorithm 
which would:
\begin{itemize}
\item[(a)] account for the most recent corrections for the muon cross sections;
\item[(b)] be of adequate and known accuracy, i.e. does not contribute an
additional systematic error which would exceed one from ``insurmountable''
uncertainties (e.g. with muon and neutrino spectra and cross sections) and 
whose value would be well known for any setting of simulation parameters;
\item[(c)] be flexible enough, i.e. could be optimized for each concrete 
purpose to desirable and well understood equilibrium between computation time 
and accuracy and be easily extended for any medium and any correction for the
cross sections of the processes in which high energy muon looses its energy;
\item[(d)] be ``transparent'', i.e. provide user with the whole set of data
related to used models for the muon cross sections;
\item[(e)] be as fast as possible.
\end{itemize}

We describe the main features of our algorithm in Sec.~\ref{sec:description}.
Sec.~\ref{sec:accuracy} gives an analysis for the algorithm accuracy. 
In Sec.~\ref{sec:simply} we report the results of investigation on the set of 
parameters which should be used to simulate the muon propagation with an 
optimum equilibrium between accuracy and computation time. 
Sec.~\ref{sec:results1} presents selected results obtained with MUM in 
comparison with ones from other muon propagation MC codes, namely PROPMU 
(Ref.~\cite{lipari}) and MUSIC (Ref.~\cite{music1}). 
Sec.~\ref{sec:conclusions1} gives general conclusions.
We also present parameterizations for muon cross sections as they are used in 
MUM in Appendix \ref{app:cs} and give proof of formula for free path between 
two muon interactions, as treated in our algorithm, in Appendix \ref{app:path}.

\section{Algorithm description}
\label{sec:description}

The basic features of the MUM algorithm are as follows.

\begin{itemize}
\item[(a)] The code does not use any preliminary computed data as an input, 
all necessary tables are prepared at the stage of initiation on the base of 
five relatively short routines, four of which return differential cross 
sections $d\sigma(E,v)/dv$ (where $E$ is muon energy, $v$ is fraction of energy
lost $v=\Delta E/E$) for bremsstrahlung, direct $e^{+}e^{-}$-pair production, 
photo-nuclear interaction and knock-on electron production, correspondingly, 
and fifth one does stopping power due to ionization $[dE(E)/dx]_{ion}$ (see 
Appendix~\ref{app:cs} for corresponding formulas). Thus, 
it is easy for any user
to correct or even entirely change the model for muon interactions as it is 
necessary. Also any material can be easily composed. Three media, namely pure 
water, ice and standard rock are available directly.
\item[(b)] We have tried to decrease the ``methodical'' part of systematic 
error which originates from finite accuracy  of numerical procedures on 
interpolation, integration, etc., down to as low level as possible, special 
attention was put on procedures which simulate free path between two sequential
muon interactions and fraction of energy lost. To combine this with the high 
speed of simulation the values for free paths, energy losses, differential and 
total cross sections along with solutions for all ordinary and integral 
equations are computed in MUM at the initiation stage, tabulated and then 
referenced when necessary with an interpolation algorithm whose accuracy has 
been carefully tested for each table by comparison with directly computed 
values to be not worse than 0.5\% 
(typically, much better).
\item[(c)] The most important parameters are changeable and can be tuned to an 
optimum combination, depending upon desirable accuracy, necessary statistics 
and restriction on the computation time for each concrete problem.
\item[(d)] The code combines algorithms for muon transportation through thick 
layers of matter down  to detector and for simulation muon interactions within 
detector sensitive volume (these algorithms have to differ from each other). 
This is important for deep underwater and under-ice Cherenkov neutrino 
telescopes (see Refs.~\cite{NT1,AMANDA1,ANTARES1,NESTOR1}), where the same 
material (water or ice) represents both a shield which absorbs atmospheric 
muons and detecting medium in which muons and shower particles resulting from 
muon interactions generate Cherenkov photons detected by phototubes. 
\item[(e)] Formally, initial muon energies up to 1 EeV can be processed by the 
MUM algorithm but uncertainties with muon cross sections which grows along with
the muon energy (especially, for photo-nuclear interaction) make one to apply 
MUM output with care at muon energies $E>$ 0.1$\div$1 PeV. 
Landau-Pomeranchuk-Migdal effect is not accounted in MUM.
\item[(d)] Besides muon transportation algorithm itself, the code includes 
number of routines which allows to obtain directly values for differential and 
total cross sections, mean free paths, energy losses and other related data for
the given set of input parameters. Sampling the atmospheric muon energies at 
the sea level according to different models for spectrum is possible with MUM, 
as well. Also several test procedures are included which provide by data 
concerning accuracy of different algorithm steps. See Sec.~\ref{sec:accuracy},
Sec.~\ref{sec:simply} and Sec.~\ref{sec:results1} for selected output of these 
procedures.
\end{itemize} 

The usual approximation for treatment the muon energy loss is applied in the 
MUM algorithm: muon interactions with comparatively large energy transfers when
fraction of energy lost $v$ exceeds some value $v_{cut}$, are accounted by 
direct simulation of $v \ge v_{cut}$  for each single interaction according to 
shape of differential cross sections (these interactions lead to ``stochastic''
energy loss or SEL) while the part of interaction with relatively small $v$ is 
treated by the approximate concept of ``continuous'' 
energy loss (CEL) using the
stopping power formula
\begin{eqnarray}
\left[
\frac{dE}{dx}(E)
\right]
_{CEL}\!&\!=&\frac{N_{A}}{A_{eff}}\rho E \sum_{j=b,p,n} \sum_{i=1}^{n}\!
\left[
k_{i}\!\!\!\int\limits_{v_{min}^{i,j}}^{v_{cut}}\!\!\frac{d\sigma_{i}^{j}(E,v)}{dv}v\,dv
\right] \nonumber\\
&\!+&
\left[
\frac{dE}{dx}(E)
\right]
_{ion} \nonumber\\
&\!-&
\frac{N_{A}}{A_{eff}}\rho E \sum_{i=1}^{n}\! 
\left[
k_{i}\!\!\!\int\limits_{v_{cut}}^{v^{i,e}_{max}}\!\!\frac{d\sigma_{i}^{e}(E,v)}{dv}v\,dv
\right]\! .
\label{CL}
\end{eqnarray} 
Here index $j$ indicates type of interaction ($j = b$ for bremsstrahlung, 
$j = p$ for direct $e^{+}e^{-}$-pair production, $j = n$ for photo-nuclear 
interaction and $j = e$ for knock-on electron production, respectively); index
$i$ runs over $n$ kinds of atoms given material consists of; 
$k_{i}=N_{i}/N_{tot}$ is fraction of $i$-th element; $N_{i}$ and $N_{tot}$ are 
number of given kind of atoms and total number of atoms, respectively, per unit
of material volume; $N_{A}$ is the Avogadro number; $\rho$ is the material 
density; $A_{eff}=N_{tot}^{-1}\sum_{i=1}^{n}(N_{i}A_{i})$ is an effective 
atomic weight for given material; $A_{i}$ is atomic weight for $i$-th element; 
$v_{min}^{i,j}$ is minimum kinematically allowed fraction of energy lost for 
$i$-th element at $j$-th process. One is forced to decompose energy losses into
two parts because simulation of all interactions with $v \ge v_{min}$ would
result in infinite computation time due to steep dependence of muon 
cross sections on $v$: they decrease with $v$ at least as 
$d\sigma(E,v)/dv \propto v^{-1}$ and for some processes are not finite at 
$v\to$ 0. Number of interactions to be simulated per unit of muon path grows,
roughly, as $N_{int} \propto v_{cut}^{-1}$ along with computation time. 
Actually, two different criteria by which the given muon interaction is 
attributed either to SEL or to CEL are available in the frame of the MUM 
algorithm. The first one (relative) has been  
described above and is applied when muon is transported down to detector
location. Second, absolute criterium, is useful when simulating muon 
interactions within an underwater or under-ice array to obtain the detector 
response with the fixed energy threshold: interaction is of  SEL type if 
$\Delta E \ge \Delta E_{cut}$ and of CEL type, if $\Delta E<\Delta E_{cut}$. 
Optionally, cross section for knock-on electron production 
$d\sigma^{e}(E,v)/dv$ can be set in MUM to zero, in which case muon 
propagation down to detector is simulated with entirely ``continuous'' 
ionization and its fluctuations are neglected but simulation of the muon 
interactions with fixed energy threshold includes knock-on electron production
in any  case. Both $v_{cut}$ and $\Delta E_{cut}$ represent parameters for MUM 
initiation procedure and can be set to any values within 
10$^{-4} \le v_{cut} \le$ 0.2 and 10 MeV $\le \Delta E_{cut} \le$ 500 MeV, 
correspondingly. The optimum value for $\Delta E_{cut}$ depends upon 
configuration of the given detector and also upon characteristics of 
algorithms which simulate the shower development, the Cherenkov photons 
generation and propagation, and detector response. All this is out of given 
article scope, therefore we discuss this parameter nowhere below except for
mentioning that Eqs. (\ref{CL}), (\ref{L}) and (\ref{sigmatot}) are used in 
algorithm with absolute treatment of the muon energy loss decomposition being 
modified by replacement $v_{cut} \to \Delta E_{cut}/E$. Influence of simplified
entirely ``continuous'' treatment of ionization and value of $v_{cut}$ upon 
simulation accuracy is analyzed in details in Sec.~\ref{sec:simply} (see also 
Ref.~\cite{MUMer}).

The principal steps of simulation are as follows. 

\begin{figure}
\hspace{4.4cm}
\mbox{\epsfig{file=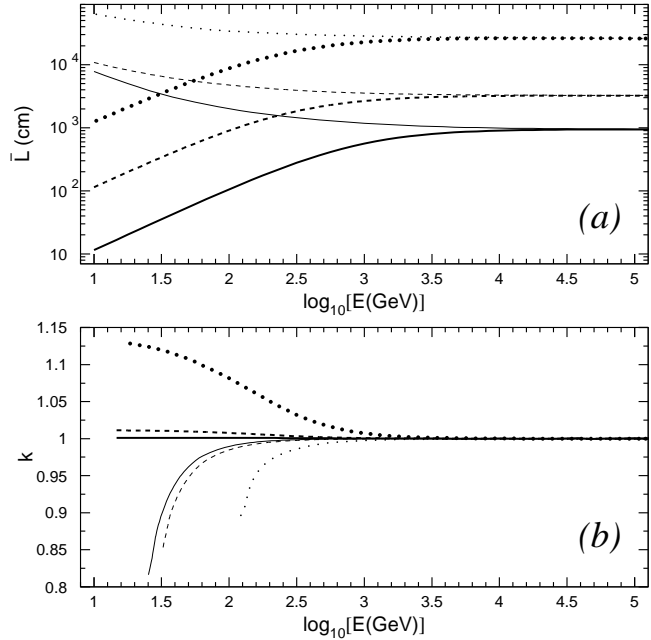,width=8.5cm}}
\protect\caption{
(a) - mean free path between two sequential muon interactions with fraction of 
energy lost $v > v_{cut}$ $\bar L(E)$ (Eq.~(\protect\ref{L})) in pure water vs.
muon energy. Two sets of curves correspond to two models of ionization. Thick 
lines are for ionization included in SEL, thin ones correspond to entirely 
``continuous'' ionization. Solid lines: $v_{cut}$ = 10$^{-4}$; dashed lines: 
$v_{cut}$ = 10$^{-3}$; dotted lines: $v_{cut}$ = 10$^{-2}$.
(b) - Function $k(E) = L_{approx}(E) / L(E)$ for $-\ln(\eta)$ = 1 (see text). 
Thickness and type of lines are of the same meaning as for plot (a). 
}
\label{fig8n}
\end{figure}

{\it I.} For muon with initial energy $E_{1}$ the free path $L$ till 
interaction with $v \ge v_{cut}$ is simulated. For this, after a random number 
$\eta$ uniformly distributed in a range from 0 to 1 has been sampled, one 
solves the following set of equations: 
\begin{equation}
\left\{
\begin{array}{rcl}
-\ln(\eta)&=&\int\limits_{E_{2}}^{E_{1}}[(dE(E)/dx)_{CEL} 
\bar L(E)]^{-1}\,dE\\
L&=&\int\limits_{E_{2}}^{E_{1}}[(dE(E)/dx)_{CEL}]^{-1}\,dE
\end{array}
\right.
\label{one}
\end{equation}
(proof of Eqs.~(\ref{one}) is given in Appendix \ref{app:path}). Here, 
$E_{2}<E_{1}$ is muon energy at the point of interaction and energy dependent 
mean free path $\bar L(E)$ between two interactions with fraction of energy 
lost $v \ge v_{cut}$ is expressed by          
\begin{equation}
\bar L(E)=\frac{A_{eff}}{\rho\,N_{A}}
\left\{
\sum_{j=b,p,n,e} \sum_{i=1}^{n} 
\!\left[
k_{i}\!\int\limits_{v_{cut}}^{v_{max}^{i,j}}\!\frac{d\sigma_{i}^{j}(E,v)}{dv}\,dv
\right]
\right\}
^{-1} ,
\label{L}
\end{equation}
where $v_{max}^{i,j}$ is maximum kinematically allowed fraction of energy lost
for $i$-th element at $j$-th kind of interaction. First equation in 
Eqs.~(\ref{one}) is solved for variable $E_{2}$, then free path $L$ can be 
found by second equation. We would like to stress that such approach allows to 
perform the accurate simulation independently on chosen value of $v_{cut}$ in 
contrast to commonly used simplification 
\begin{equation}
L_{approx} = -\bar L(E_{1}) \ln (\eta) ,
\label{Lap}
\end{equation}
which neglects dependence $\bar L(E)$ upon energy and, consequently is 
{\it (a)} the less accurate the larger $v_{cut}$ is and {\it (b)} produces the 
error of different signs for the cases when ionization is included in SEL or its
fluctuations are neglected. It is illustrated by two plots in  Fig.~\ref{fig8n}.
Upper plot shows function $\bar L(E)$ for pure water. Two sets of curves are 
presented for two models of ionization. Each set includes dependencies for 3 
values of $v_{cut}$: 10$^{-4}$, 10$^{-3}$ and 10$^{-2}$. In fact, $\bar L(E)$ 
is almost a constant at $E >$ 5 TeV but changes steeply at lower energies. It 
increases with decrease of energy if ionization is entirely ``continuous'' and,
on the contrary, it decreases if ionization is included in SEL. Thus, 
simulating
free path by Eq.~(\ref{Lap}) one overestimates it (and consequently 
underestimates energy loss) in case when ionization is included in SEL and, on 
the contrary one underestimates free path and overestimates energy loss in case
of completely ``continuous'' ionization. Lower plot in Fig.~\ref{fig8n} shows 
resulting error in the value of simulated free path if $-\ln(\eta)$ = 1 (for 
larger $-\ln(\eta)$ the effect is more significant). The set of curves 
represent dependencies $k(E) = L_{approx}(E) / L(E)$ with $L(E)$ computed by 
Eqs.~(\ref{one}) and $L_{approx}(E)$ computed by Eq.~(\ref{Lap}). With 
ionization included in SEL overestimation for free path is less than 1\% 
at $v_{cut} \le$  10$^{-3}$ but reaches $\sim$15\% 
at  $v_{cut}$ = 10$^{-2}$ which leads to 1$\div$2\% 
underestimation of total energy loss below muon energy 1 TeV. In case with 
``continuous'' ionization the effect is of opposite sign and again is more 
significant for large $v_{cut}$.

{\it II.} After free path $L$ and muon energy $E_{2}$ have been found from 
Eqs.~(\ref{one}), the type of interaction is simulated according to proportion 
between total cross sections of different processes:
\begin{equation}
\sigma^{b}\;:\;\sigma^{p}:\;\sigma^{n}\;:\;\sigma^{e}
\label{type}
\end{equation}
which are computed as:
\begin{equation}
\sigma^{j}=
\sum_{i=1}^{n}
\!\left[
k_{i}\!\int\limits_{v_{cut}}^{v_{max}^{i,j}}\!\frac{d\sigma_{i}^{j}(E_{2},v)}{dv}\,dv
\right].
\label{sigmatot}
\end{equation}
                                              
{\it III.} Fraction of energy lost $v$ is simulated according to shape of 
differential cross section for given process $j$:
\begin{equation}
\frac{d \sigma^{j}}{dv}(E_{2},v)=\sum_{i=1}^{n}k_{i}
\frac{d \sigma_{i}^{j}}{dv}(E_{2},v)
\label{cscomp}
\end{equation}
and new muon energy $E_{1}^{'}=E_{2} \cdot (1 - v)$ is determined.

{\it IV.} Steps {\it I--III} are repeated sequentially until muon either 
reaches the  level of observation or stops. Muon are considered as stopped
as soon as its energy decreases down to 0.16 GeV which corresponds to the 
Cherenkov threshold for muon in pure water.  

\section{Algorithm accuracy}
\label{sec:accuracy}

\begin{figure}
\hspace{3.6cm}
\mbox{\epsfig{file=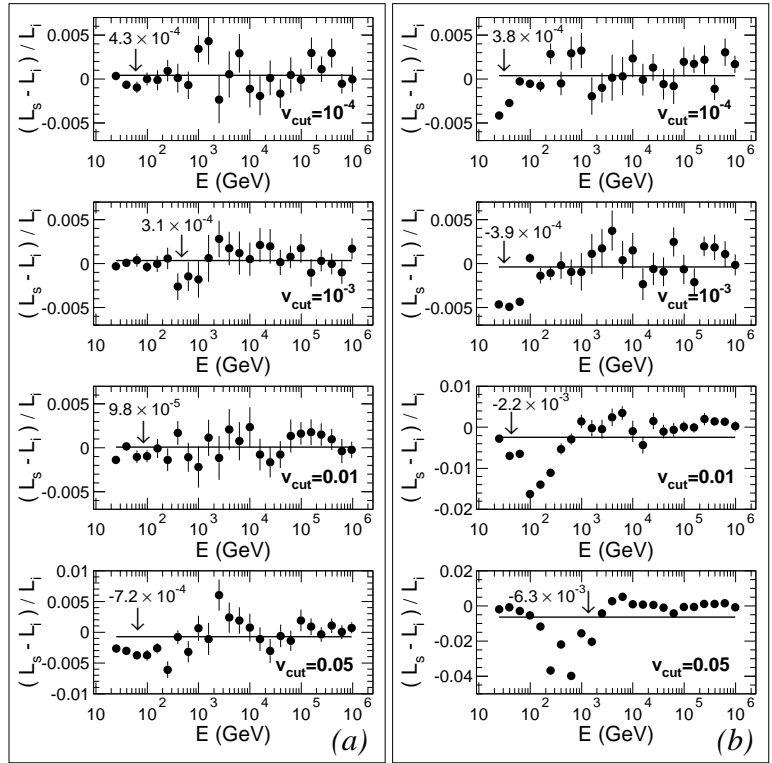,width=10.2cm}}
\protect\caption{
Relative difference $(L_{s}-L_{i})/L_{i}$ between ``simulated'' $L_{s}$ 
(Eq.~(\ref{Ls})) and ``integrated'' $L_{i}$ (Eqs. (\ref{Li}), (\ref{Li1})) 
total muon 
energy loss in the pure water. Horizontal solid line on each plot shows 
averaged over 24 tested muon energies value for $(L_{s}-L_{i})/L_{i}$ which, 
additionally, is given at the upper left corner by the figure. 
Statistical error at 1$\sigma$-level is shown at each point. 
(a) - ionization is included in SEL; (b) - ionization is entirely
``continuous''.
}
\label{test1}
\end{figure}

\begin{figure}
\hspace{3.6cm}
\mbox{\epsfig{file=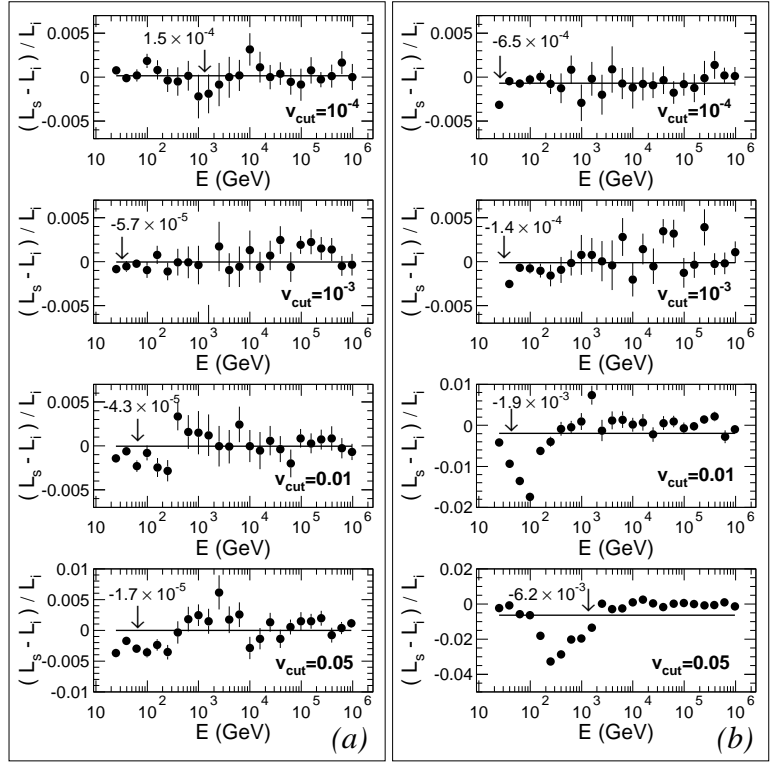,width=10.2cm}}
\protect\caption{
The same as in Fig.~\protect\ref{test1} for standard rock 
($\rho$ = 2.65 g cm$^{-3}$, $A$ = 22, $Z$ = 11).
}
\label{test2}
\end{figure}

As described in Sec.~\ref{sec:description} the MUM code (as well as any muon MC
propagation algorithm) consists of the set of procedures on numerical solution 
of equations, interpolation and integration. All these procedures are of finite
accuracy and, consequently, the incoming model for muon energy loss is somewhat
corrupted by them. Thus, resulting energy loss as it {\it simulated} by a code
are not the same as energy loss as it can be calculated by {\it integration} 
the differential cross sections which are at the input of the same code. The 
difference between simulated and calculated energy loss contains errors which 
are contributed by each step of simulation algorithm and thus, is a good 
quantitative criterium for its {\it inner accuracy}, whose contribution to the 
resulting error must not exceed one which comes, e.g. from uncertainties with 
muons cross sections and medium composition. Therefore to demonstrate accuracy
of presented algorithm we have chosen just data on the relative difference 
$(L_{s}-L_{i})/L_{i}$ between simulated $L_{s}$ and integrated $L_{i}$ total 
muon energy loss as was obtained with MUM for the pure water (Fig.~\ref{test1})
and standard rock ($\rho$ = 2.65 g cm$^{-3}$, $A$ = 22, $Z$ = 11, 
Fig.~\ref{test2}).
Inner accuracy is presented in the figure as a function of muon energy 
for several values of $v_{cut}$ and two models of ionization energy loss. 
Values of $L_{s}$ were obtained as follows. For each muon energy $E_{1}$ a 
distance $D$ was chosen and propagation of $N$ = 4$\times$10$^{6}$ muons over 
this distance was simulated. The condition $N \cdot D \gg \bar L(E_1)$ must be 
obeyed to obtain statistically significant result but, at the same time, $D$ 
should be short enough to be passed by muons without decrease their energy down
to zero, which practically leads to $D =$ 0.5 m$\div$300 m depending upon muon 
energy, value of $v_{cut}$ and kind of medium. For each $i$-th muon its final 
energy $E_{2}^{i}$ was fixed and then $L_{s}$ was calculated as
\begin{equation}
L_{s}=
\frac{1}{D}
\left(
E_{1}-\frac{1}{N}\sum_{i=1}^{N}E_{2}^{i}
\right).
\label{Ls}
\end{equation} 
$L_{i}$ was computed as 
\begin{equation}
L_{i}=\frac{1}{D}(E_{1}-E_{2}),
\label{Li}
\end{equation}
\noindent
where $E_{2}$ was found as a solution of integral equation for the muon range:
\begin{equation}
D=\!\int\limits_{E_{2}}^{E_{1}}
\left\{
\frac{N_{A}}{A_{eff}}\rho E \sum_{j=b,p,n} \sum_{i=1}^{n}\!
\left[
k_{i}\!\!\!\int\limits_{ v_{min}^{i,j}}^{v_{max}^{i,j}}\!\!\frac{d\sigma_{i}^{j}(E,v)}{dv}v\,dv
\right]+
\left[
\frac{dE}{dx}(E)
\right]
_{ion}
\right\}
^{-1}\!\!\!\!\!dE .
\label{Li1}
\end{equation} 
Horizontal solid line on each plot shows the value for $(L_{s}-L_{i})/L_{i}$ 
averaged over 24 tested muon energies which, additionally, is given at upper 
left corner by a figure. Fig.~\ref{test1}(a) and Fig.~\ref{test2}(a) indicate 
an excellent inner accuracy of the MUM algorithm with ionization included in 
SEL both for water and standard rock. Up to $v_{cut}$ = 5$\times$10$^{-2}$ all 
points are within 0.6\%-deviations 
which, besides, are of both signs, so averaged accuracy remains better than 
10$^{-3}$. Fig.~\ref{test1}(b) and Fig.~\ref{test2}(b) (which correspond to 
simplified completely ``continuous'' ionization) shows somewhat worse accuracy 
of the algorithm which falls down when $v_{cut}$ increases. Accuracy was found 
to be within 1\%
(with except for few points around muon energy $E$ = 100 GeV) up to 
$v_{cut}$ = 10$^{-2}$ with averaged accuracy within 2$\times$10$^{-3}$. This 
last value may be used as a conservative evaluation of inner accuracy for the 
MUM algorithm. Statistically significant likeness of plots obtained for water 
and standard rock can be seen.

Thus, we conclude that assuming an optimistic evaluation of 1\% 
for uncertainties in muon cross sections (Refs.~\cite{rhode1,kp}) the inner 
inaccuracy of MUM does not exceed them for any $v_{cut} \le$ 5$\times$10$^{-2}$
if ionization is included in SEL and for any $v_{cut} \le$ 10$^{-2}$ if 
ionization is treated as entirely ``continuous process'', independently upon 
material.

\section{The optimum setting of simulation parameters}
\label{sec:simply}

As was described in Sec.~\ref{sec:description} $v_{cut}$ is a parameter in the 
MUM algorithm and can be set optionally to different values. The larger is 
$v_{cut}$ the higher is the speed of simulation, because the less muon 
interactions have to be simulated per unit of the muon path. But, on the other 
hand, too large value of $v_{cut}$ leads to the lost of accuracy since some 
essential part of fluctuations in the muon energy losses comes out of direct 
simulation. Thus, the question is {\it how large value of $v_{cut}$ may be 
chosen to keep result within desirable accuracy}? Also different models for 
ionization can be used: it can be  optionally either treated as completely 
``continuous'' process or included in SEL. Small energy transfers strongly 
dominate at knock-on electron production 
($d\sigma^{e}(E,v)/dv \propto v^{-2}$), so this process is almost 
non-stochastic and it seems to be reasonable to exclude knock-on electrons from 
simulation procedure  when simulating SEL which saves computation time 
noticeable. {\it How much does it affect the result of simulation}? Influence 
of these factors on simulated result had been discussed in literature (see, 
e.g. Refs.~\cite{N94,lipari,music1,lagutin1}) but in our opinion more detailed 
analysis was lacking. Therefore we have undertaken our own investigation which 
is reported in this Section. For that we performed several sets of simulations 
both for propagation of mono-energetic muon beams and atmospheric muons sampled
by sea level spectrum (in the later case we limited ourselves by simulation 
only vertical muons) through pure water down to depths from $D =$ 1 km to 
$D =$ 40 km. Of course, distances of more than several kilometers for vertical 
muons do not concern any real detector but simulations for large depths allow 
us to study general regularities which correspond, e.g. to nearly horizontal 
directions. Several runs were done for standard rock, as well. We tested 
different settings of parameters which were as follows.

\begin{itemize}
\item[(a)] $v_{cut}$, which changed within a range of 
10$^{-4} \le v_{cut} \le$ 0.2. Inner accuracy of the MUM code becomes somewhat 
worse at $v_{cut} \ge$ 5$\times$10$^{-2}$, especially if fluctuations in 
ionization are not simulated (Sec.~\ref{sec:accuracy}) therefore results for 
$v_{cut}$ = 0.1 and $v_{cut}$ = 0.2 are presented here only to illustrate some 
general qualitative regularities.
\item[(b)] Model for ionization.
\item[(c)] Parameterization for vertical sea level atmospheric muon spectrum. 
Two spectra were tested, namely one proposed in Ref.~\cite{bks1} (basic):
\begin{equation}
\frac{dN}{dE}=\frac{0.175\:E^{-2.72}}{cm^{2}\:s\:sr\:GeV}
\left(
\frac{1}{\displaystyle
  1 + E/103 GeV} +
\frac{0.037}{\displaystyle
  1 + E/810 GeV}
\right),
\label{bknsspec}
\end{equation}
and the Gaisser spectrum (Ref.~\cite{gaisser}):
\begin{equation}
\frac{dN}{dE}=\frac{0.14\:E^{-2.7}}{cm^{2}\:s\:sr\:GeV}
\left(
\frac{1}{\displaystyle
  1 + E/104.6 GeV} +
\frac{0.054}{\displaystyle
  1 + E/772.7 GeV}
\right).
\label{gaisspec}
\end{equation}
\item[(d)] Parameterization for total cross section for absorption of a real
photon by a nucleon at photo-nuclear interaction $\sigma_{\gamma N}$ which was 
treated both according to the Bezrukov-Bugaev parameterization proposed in 
Ref.~\cite{phnubb} (basic) and the ZEUS parameterization (Ref.~\cite{ZEUS})
(see. Appendix \ref{app:phnu} for formulas).
\item[(e)] A factor $k_{\sigma}$ which all muon cross sections along with 
stopping power due to ionization were multiplied by to test influence of 
uncertainties in muon cross sections (and, consequently, in energy losses) upon
result. We applied $k_{\sigma}$ = 1.0 as a basic value but set also 
$k_{\sigma}$ = 0.99 and $k_{\sigma}$ = 1.01, which corresponds to decrease 
and increase of total energy loss by 1\%, 
respectively. Note that it is an ``optimistic'' evaluation, the real 
accuracy of existing parameterization for muon cross sections is worse (see 
Refs.~\cite{rhode1,kp}).
\end{itemize} 

For each run we fixed the muon spectra at final and several interim depths.
The differences between obtained spectra were a point of investigation.

\begin{figure}
\hspace{1.4cm}
\mbox{\epsfig{file=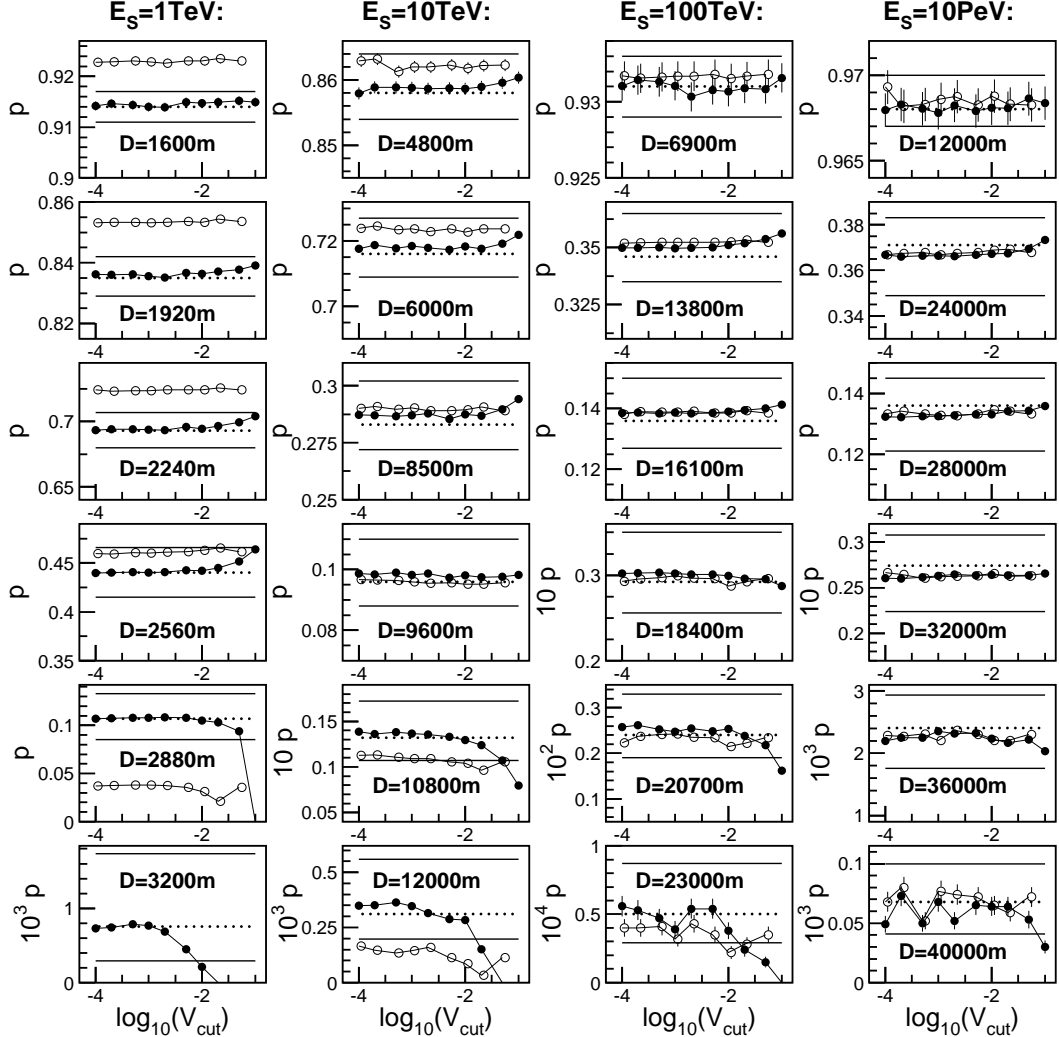,width=14.0cm}}
\protect\caption{
Survival probabilities $p=N_{D}/N_{s}$ (where $N_{s}$ = 10$^{6}$ is initial
number of muons in the beam and $N_{D}$ is number of muons which have survived 
after propagation down to slant depth $D$ in pure water) vs. $v_{cut}$. Values 
of $p$ were obtained as a result of simulation with MUM for mono-energetic muon
beams with initial energies $E_{s}$ = 1 Tev (1st column of plots), 10 TeV (2nd 
column), 100 TeV (3rd column) and 10 PeV (4th column). Each column contains six
plots which correspond to six slant depths $D$ (which differs for different 
$E_{s}$). Closed circles represent survival probabilities which were simulated 
with ionization energy losses included in SEL, open ones correspond to 
computation with completely ``continuous'' model of ionization. Two horizontal 
solid lines on each plot show the value for survival probability computed with 
all muon cross sections multiplied by a factor $k_{\sigma}$ = 1.01 (lower line)
and $k_{\sigma}$ = 0.99 (upper line) for $v_{cut}$ = 10$^{-4}$. Horizontal 
dotted lines correspond to $v_{cut}$ = 10$^{-4}$ and cross section for 
absorption of a real photon at photo-nuclear interaction parameterized 
according to Ref.~\protect\cite{ZEUS} instead of Ref.~\protect\cite{phnubb}
which is basic in MUM. Note different scales at Y-axis.
}
\label{fig1}
\end{figure}

At the first set of simulations we propagated mono-energetic muon beams of
4 fixed initial energies $E_{s}$ = 1 TeV, 10 TeV, 100 TeV and 10 PeV down
to slant depths $D$ = 3.2 km, 12 km, 23 km and 40 km, respectively, through 
pure water. The value $D$ for each initial muon energy was chosen so that 
majority of muons had been stopped after propagation the given distance. This
allows to track differences in simulated results obtained with different 
settings of parameters for all segments of muon beam path. In each case 
propagation of 10$^{6}$ muons was simulated. Fig.~\ref{fig1} shows resulting 
survival probabilities $p=N_{D}/N_{s}$ (where $N_{s}$ = 10$^{6}$ is initial 
number of muons and $N_{D}$ is number of muons which have survived after 
propagation down to the slant depth $D$) vs. $v_{cut}$ for final and five 
interim values of $D$. Two curves are given on each plot for two models of 
ionization. Also results for $k_{\sigma}$ = 1.00 $\pm$ 0.01 and for 
$\sigma_{\gamma N}$ parameterization according to Ref.~(\cite{ZEUS}) are 
presented as simulated with the most accurate value $v_{cut}$ = 10$^{-4}$.

The following conclusions can be done.
\begin{itemize}
\item[(a)] In most cases except for some plots of the lower row and 
the left column in Fig.~\ref{fig1} (which corresponds to low survival 
probabilities and low muon initial energies, respectively) uncertainty in 
our knowledge of muon cross sections gives the principal effect which 
essentially exceeds ones from other tested parameters.
\item[(b)] The difference between survival probabilities for two models of 
ionization is the less appreciable the larger muon energy is. It is quite 
understandable because at muon energies $E<$ 1 TeV ionization represents 
the great bulk of total energy loss, and vice versa, it becomes minor at 
$E>$ 1 TeV. Thus, contribution which is given by ionization at higher
energies is small and, the more, its fluctuations do not play an important
role. For muons with initial energies $E \gg$ 1 TeV fluctuations in
ionization become important only at very last part of muon path and 
``are not in time'' to produce some noticeable effect.
\item[(c)] Generally, parameterizations for $\sigma_{\gamma N}$ as proposed
in Refs.~\cite{phnubb,ZEUS} do not show a noticeable difference in terms of 
survival probabilities, in most cases it is within statistical error or exceeds
it only slightly.
\item[(d)] Increase of $v_{cut}$ gives effect of both signs in survival 
probabilities: function $p(v_{cut})$ grows at the beginning of muon path
and falls at the last part. The same ``both-sign'' dependencies are observed
for ionization model. 
\item[(e)] For $v_{cut} \le$ 0.05 there is almost no dependence of
survival probability on $v_{cut}$ except for very last part
of muon path where survival probability becomes small. Generally,
dependence $p(v_{cut})$ is the less strong the larger initial muon
energy is.
\end{itemize}

\begin{figure}
\hspace{3.5cm}
\mbox{\epsfig{file=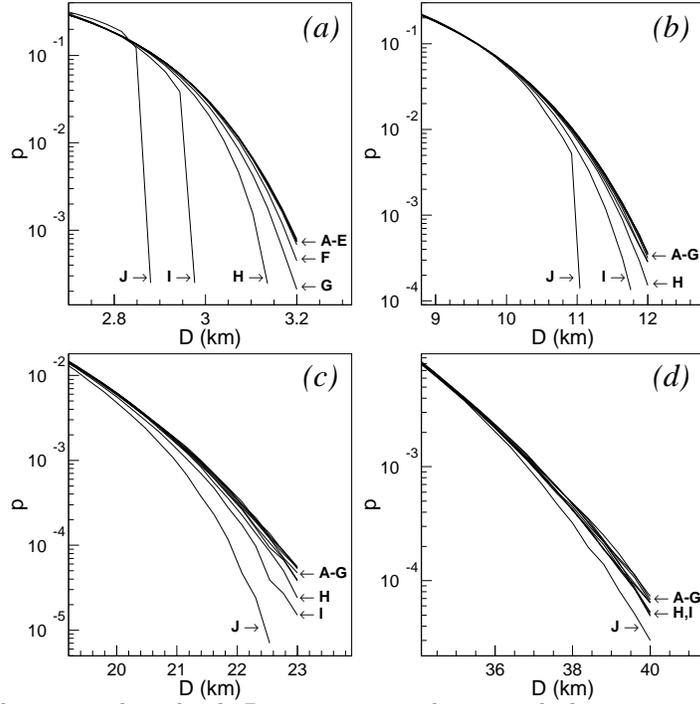,width=9.3cm}}
\caption{Survival probability $p$ vs. slant depth $D$ in pure water down to 
which propagation of mono-energetic muon beam with initial energy 
$E_{s}$ = 1 TeV (a), 10 TeV (b), 100 TeV (c) and 10 PeV (d) is simulated. On 
each plot 10 lettered curves which correspond to different values of $v_{cut}$
are shown. Meaning of letters is as follows: 
A - $v_{cut}$ = \mbox{10$^{-4}$}, 
B - $v_{cut}$ = \mbox{2$\times$10$^{-4}$}, 
C - $v_{cut}$ = \mbox{5$\times$10$^{-4}$}, 
D - $v_{cut}$ = \mbox{10$^{-3}$}, 
E - $v_{cut}$ = \mbox{2$\times$10$^{-3}$}, 
F - $v_{cut}$ = \mbox{5$\times$10$^{-3}$}, 
G - $v_{cut}$ = \mbox{10$^{-2}$}, 
H - $v_{cut}$ = \mbox{2$\times$10$^{-2}$}, 
I - $v_{cut}$ = \mbox{5$\times$10$^{-2}$}, 
J - $v_{cut}$ = \mbox{10$^{-1}$}. 
This figure displays results which were obtained by simulation with 
ionization losses included in SEL. Statistical errors (which cause some
un-smoothness of curves at small $p$) are not shown.
}
\label{fig2}
\end{figure}

\begin{figure}
\hspace{3.5cm}
\mbox{\epsfig{file=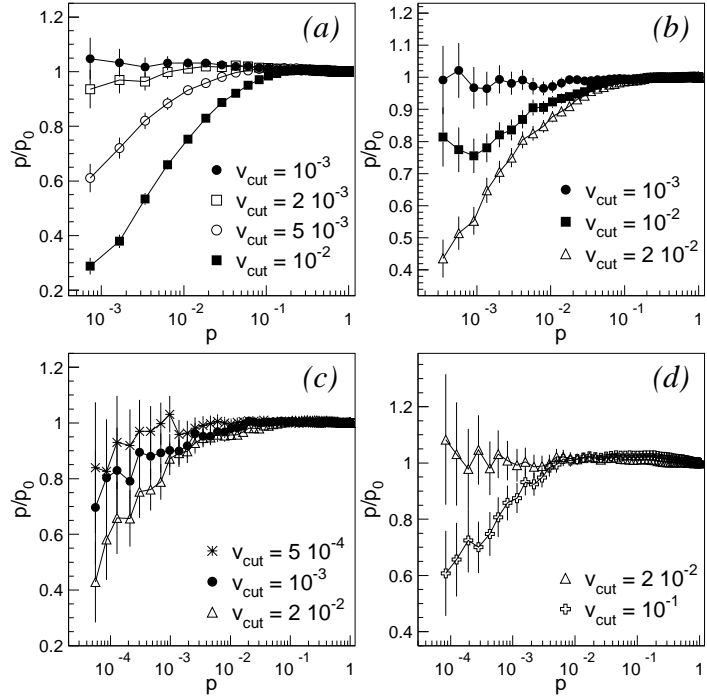,width=9.3cm}}
\caption{Relation $p/p_{0}$ vs. $p$. $p$ is survival probability for muon s
of initial energy $E_{s}$ = 1 TeV (a), 10 TeV (b), 100 TeV (c) and 10 PeV (d) 
at propagation through pure water with ionization included in SEL as simulated 
for different values of $v_{cut}$; $p_{0}$ is survival probability simulated 
under the same conditions for $v_{cut}$ = 10$^{-4}$. Difference in $p/p_{0}$ 
becomes noticeable only at $p < $10$^{-1}$, i.e. at the very last part of muon 
beam path.
}
\label{fig3}
\end{figure}

\noindent
The last item is illustrated complementary by Fig.~\ref{fig2} and
Fig.~\ref{fig3} which show that for all initial energies $E_{s}$ simulated 
survival probability does not depend, in fact, on $v_{cut}$ until 90\%
(for $E_{s}$ = 1 TeV) to 99.5\%
(for $E_{s}$ = 10 PeV) muons have been stopped. 

\begin{figure}
\hspace{1.1cm}
\mbox{\epsfig{file=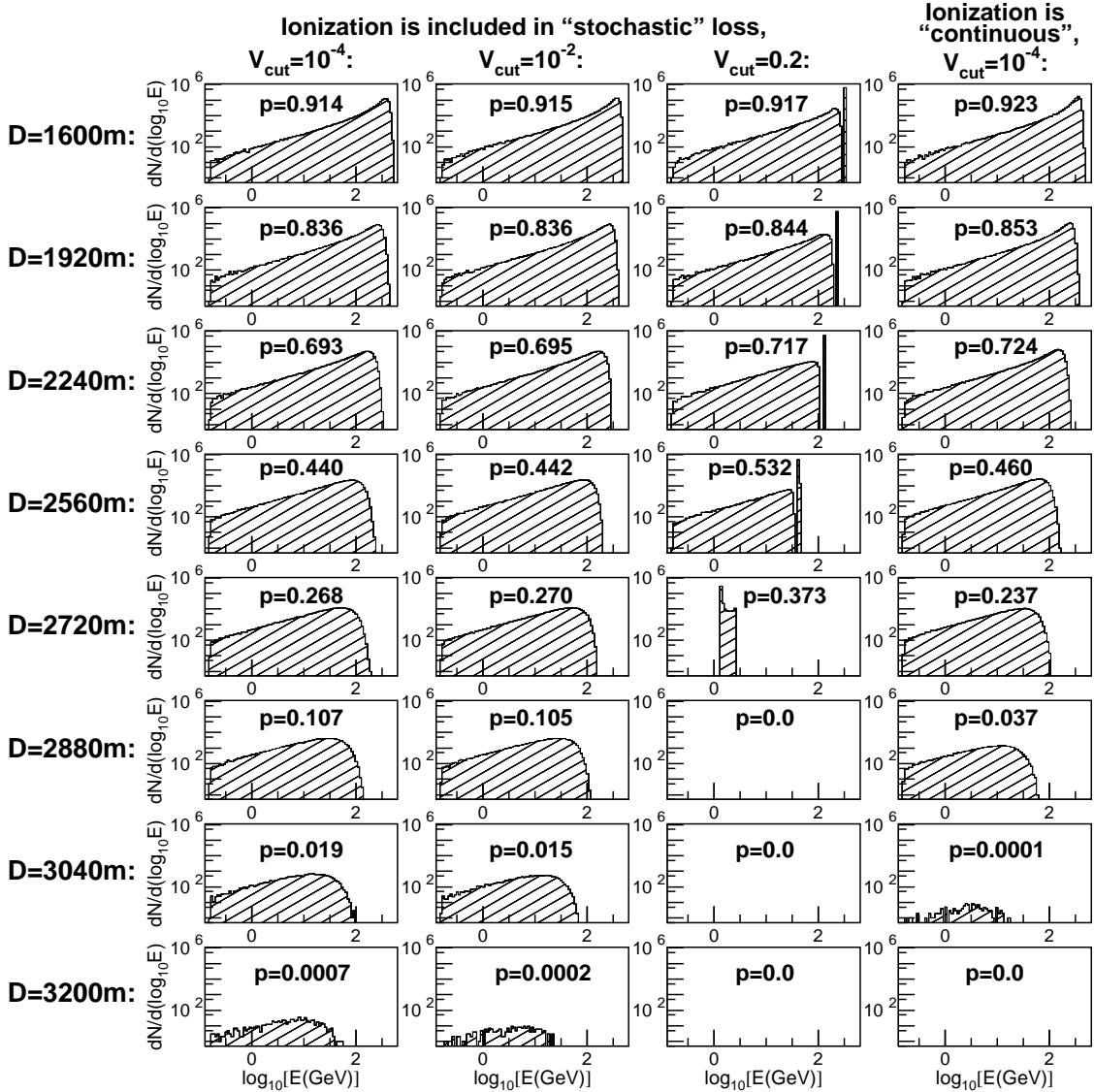,width=15.0cm}}
\caption{Muon spectra  resulting from 10$^{6}$ muons with initial energy 
$E_{s}$ = 1 TeV as simulated with four models for different slant depths $D$ in
pure water.  The first three columns represent spectra obtained with ionization
included in SEL for $v_{cut}$ = 10$^{-4}$ (1st column), 10$^{-2}$ (2nd column),
and 0.2 (3rd column). 4th column contains spectra obtained for entirely 
``continuous'' ionization and $v_{cut}$ = 10$^{-4}$. On each plot value of 
survival probability $p$ is indicated without statistical error which does not 
exceed 1\%.
}
\label{fig4n}
\end{figure}

It was shown above {\it what is result} of simulations with different models of
ionization and values of $v_{cut}$. It was a special point of interest for us 
to track {\it how and why} does it influence upon behavior of survival 
probability. Fig.~\ref{fig4n} shows how muon spectrum resulting from 
mono-energetic muon beam with initial energy $E_{s}$ = 1 TeV transforms when 
its propagation being simulated through pure water down to the slant depth of 
3.2 km. Results for four settings of parameters are presented by four columns 
of plots. The first three columns represent spectra obtained with ionization 
included in SEL for $v_{cut}$ = 10$^{-4}$, 10$^{-2}$ and 0.2. 4th column 
contains spectra simulated with entirely ``continuous'' ionization and 
$v_{cut}$ = 10$^{-4}$. The spectra grouped into the first column represent
the most accurate tuning both for $v_{cut}$ and ionization model. The first 
three columns demonstrate that compactness of spectra at the same slant depth 
is the higher the more value of $v_{cut}$ is. Put your attention to the right 
edge of spectra which shifts toward low energies when $v_{cut}$ increases (it 
is the most noticeably for $v_{cut}$ = 0.2). The reason is that at any slant 
depth energy of the most energetic muons in simulated beam is determined by 
CEL. These muons due to statistical fluctuations did not undergo interactions 
with $v \ge v_{cut}$ and, consequently, lost energy only by CEL which grows
when $v_{cut}$ increases. That is why the maximum energy in simulated muon beam
is lower for large values of $v_{cut}$. Fraction of muons which did not undergo
an ``catastrophic'' act with $v \ge v_{cut}$ till given slant depth grows with 
increase of $v_{cut}$ because free path between two sequential interactions 
with $v \ge v_{cut}$ grows approximately as $\bar L \propto v_{cut}$. It leads,
in particular, to distinctly visible separated picks in spectra for 
$v_{cut}$ = 0.2 consisted just of muons which lost energy only by CEL. Also 
some deficit of low energy muons appears if one sets $v_{cut}$ to a large 
value. In this case left edge of spectrum is provided only with muons which 
interacted with large fraction of energy lost while for smaller $v_{cut}$ an 
additional fraction of muons comes here. As a result simulated spectrum of 
initial mono-energetic muons at given slant depth is more narrow if $v_{cut}$ 
is large and, on the contrary, more wide if $v_{cut}$ is small. 

Now it is easy to understand how value of $v_{cut}$ influences on simulated 
survival probabilities. When simulated muon beam goes through a medium loosing
energy both in CEL and SEL processes, its spectrum is constantly shifting to 
the left (energy decreases). For $v_{cut}$ = 10$^{-4}$ the left part of 
spectrum reaches $E$ = 0 at a smaller slant depth comparing with larger 
$v_{cut}$ and survival probability starts to decrease. At the same slant depth 
survival probability for $v_{cut}$ = 10$^{-2}$ and $v_{cut}$ = 0.2 is still 
equal to 1. Thus, 
for the first part of path the survival probability is always 
larger for large $v_{cut}$. At some slant depth (which is equal to $\sim$2.8 km
in the given case) compactness of spectra simulated with large $v_{cut}$ starts 
to play an opposite role. Due to more powerful CEL muons stop faster comparing 
with accurate simulation. So at the final part of the beam path simulated 
survival probability for large $v_{cut}$ decreases faster comparing with 
accurate simulation and, for instance, for $v_{cut}$ = 0.2 the rest of muon 
beam which reaches the slant depth $D$ = 2.72 km (37\%
of initial number of muons) completely vanishes within the next 30 m of path, 
while some fraction of muons simulated with $v_{cut}$ = 10$^{-4}$ (0.07\%)
escapes down to the slant depth of $D$ = 3.2 km. Qualitatively the same effect 
leads to the same consequences if one treats ionization as completely 
``continuous''process. Again, spectra becomes more narrow since fluctuations in 
ionization do not work and, as a consequence, survival probability becomes 
significantly higher comparing with simulation with accurate treatment of 
ionization at the beginning of muon beam path and falls down essentially faster
at the final part of path. 

\begin{figure}
\hspace{1.8cm}
\mbox{\epsfig{file=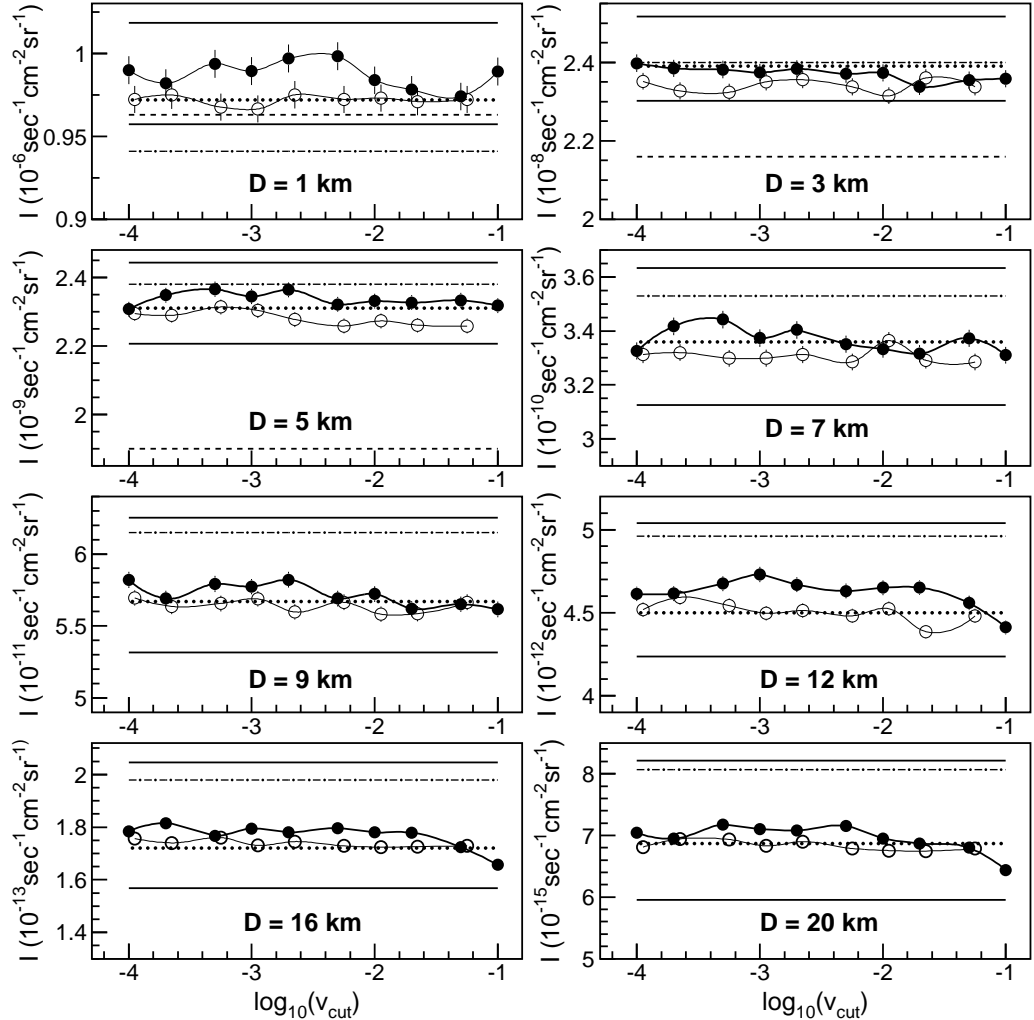,width=13.6cm}}
\protect\caption{
Intensity of vertical atmospheric muon flux $I$ at different depths $D$ of pure
water vs. $v_{cut}$ as obtained by simulation with muons sampled according to 
sea level spectrum from Ref.~\protect\cite{bks1} 
(Eq.~(\protect\ref{bknsspec})). Closed circles: ionization is included in SEL;
open circles: ionization is completely ``continuous''. Two horizontal solid
lines on each plot show value for survival probability simulated with all muon 
cross sections multiplied by a factor  $k_{\sigma}$ = 1.01 (lower line) and 
$k_{\sigma}$ = 0.99 (upper line) for $v_{cut}$ = 10$^{-4}$. Dashed lines on 
plots for $D \le$ 5 km correspond to intensity which was calculated for all 
energy losses treated as ``continuous''. Dash-dotted lines show intensity of 
vertical muon flux simulated with ionization included in SEL, 
$v_{cut}$ = 10$^{-4}$ and muons sampled according to the Gaisser sea level 
spectrum
(Ref. \protect\cite{gaisser}, Eq.~(\protect\ref{gaisspec})). Horizontal dotted 
lines correspond to $v_{cut}$ = 10$^{-4}$ and cross section for absorption of a
real photon at photo-nuclear interaction parameterized according to 
Ref. \protect\cite{ZEUS} instead of parameterization proposed in 
Ref.~\protect\cite{phnubb} which is basic in MUM. 
}
\label{fig5n}
\end{figure}

Results presented above showed the significant influence which both model of 
ionization and value of $v_{cut}$ have over survival probability for 
mono-energetic muon beam. But for practical purposes the more important is 
{\it how these factors do work for real atmospheric muons with a power 
spectrum}? In Fig.~\ref{fig5n} we present intensity of vertical atmospheric 
muon flux $I$ at different depths of pure water $D$ from 1 km to 20 km vs. 
$v_{cut}$ as simulated with muons sampled according to sea level spectrum 
Eq.~(\ref{bknsspec}). Simulation continued until 10$^{4}$ muons reached given 
depth. Curves for two models of ionization are shown for each depth along 
with results for $k_{\sigma}$ = 1.00 $\pm$ 0.01 at $v_{cut}$ = 10$^{-4}$, 
parameterization for $\sigma_{\gamma N}$ from Ref.~\cite{ZEUS} at 
$v_{cut}$ = 10$^{-4}$, sea level muon spectrum Eq.~(\ref{gaisspec}) at 
$v_{cut}$ = 10$^{-4}$ and all energy losses treated entirely as CEL (for depths
$D \le$ 5 km only). General conclusions for case with atmospheric muons are 
qualitatively the same as observed for mono-energetic muon beams but 
quantitatively the influence of $v_{cut}$ and model of ionization energy losses
on the resulting muon flux at large depths is much weaker. One can conclude the
following:   
\begin{itemize}
\item[(a)] Except for case $D$ = 1 km computed muon flux is strongly affected by
accounting for fluctuations in energy losses: muon flux intensity simulated 
with non-stochastic model of energy loss is less comparing with stochastic 
model by 10 \%  
at 3 km w.e. and by 20\%
at 5 km w.e.. At the depth of 20 km of pure water vertical muon flux computed 
with ignorance of fluctuations is only 10 \% 
of simulated flux.
\item[(b)] Like a case with mono-energetic beams 1\%-uncertainty 
in muon cross sections plays the principal role for resulting error in simulated
muon depth intensity. This error has a tendency to grow with depth from 
$\pm$2.5\% 
at depth of 1 km w.e. to $\sim \pm$15\% 
at 20 km w.e.. But a particular case of this uncertainty, namely difference 
between parameterizations for $\sigma_{\gamma N}$ from 
Refs.~\cite{phnubb,ZEUS}, does not lead to a significant difference in 
resulting intensity.
\item[(c)] Difference between muon spectra Eq.~(\ref{bknsspec}) and 
Eq.~(\ref{gaisspec}) leads to uncertainty from -4\% 
($D$ = 1 km) to 16\% 
($D$ = 20 km). 
\item[(d)] Error which appears due to simplified, entirely ``continuous''
ionization lies, commonly, at the level of 2$\div$3\%.
\item[(e)] Dependence of simulated muon flux intensity upon $v_{cut}$ is the
most weak one comparing with other studied error sources. Function $I(v_{cut})$
is almost a constant at $v_{cut} \le$ 0.05 and changes in a range 
$\pm$1$\div$2\%
which is very close to statistical error. Up to $v_{cut}$ = 0.1 the contributed
error is less than one which comes from $\pm$1\%--uncertainty 
with the muon cross sections. Also no statistically significant
influence of $v_{cut}$ upon the shape of differential atmospheric muon spectra 
was observed at all tested depths for 10$^{-4} \le v_{cut} \le$ 0.2 for both 
models of ionization energy loss. 
\end{itemize}

\begin{figure}
\hspace{5.0cm}
\mbox{\epsfig{file=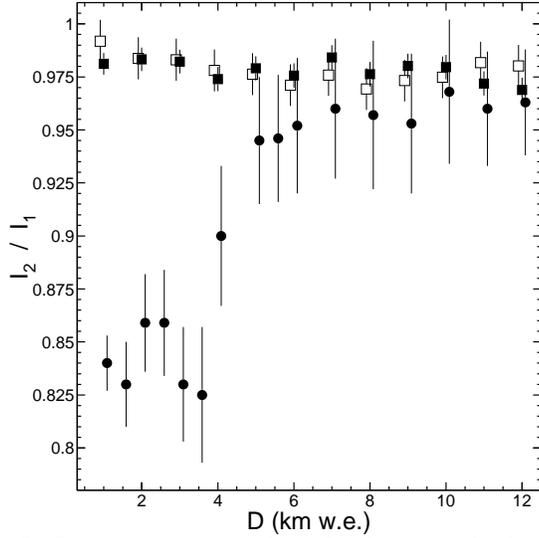,width=7.1cm}}
\protect\caption{
Dependencies for relation $I_{2}/I_{1}$ vs. water equivalent depth in 
standard rock as computed in this work (closed squares), 
in Ref.~\protect\cite{music1} (open squares) and in 
Ref.~\protect\cite{lagutin1} 
(closed circles). $I_{1}$ is depth intensity for vertical atmospheric muon flux 
simulated with ionization included in SEL, $I_{2}$ is one simulated with
entirely ``continuous'' ionization. Data for this work are obtained
for sea level atmospheric muon spectrum from Ref.~\protect\cite{bks1} 
(Eq.~(\protect\ref{bknsspec})) and $v_{cut}$ = 10$^{-3}$; data from
Ref.~\protect\cite{music1} represent result of simulation for sea level spectrum
from Ref.~\protect\cite{gaisser} (Eq.~(\protect\ref{gaisspec})) and 
$v_{cut}$ = 10$^{-3}$; data from Ref.~\protect\cite{lagutin1} were
simulated with spectrum from Ref.~\protect\cite{volkova} with
``small  transfer grouping'' technique.
}
\label{fig9n}
\end{figure}

\begin{figure}
\hspace{5.0cm}
\mbox{\epsfig{file=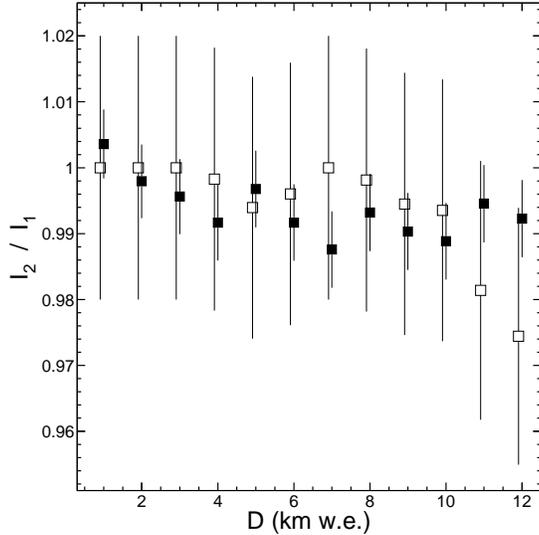,width=7.1cm}}
\protect\caption{
Dependencies for relation $I_{2}/I_{1}$ vs. water equivalent depth in 
standard rock as computed in this work (closed squares) and in 
Ref.~\protect\cite{music1} (open squares). $I_{1}$ is depth intensity for 
vertical atmospheric muon flux simulated with entirely ``continuous'' 
ionization and
$v_{cut}$ = 10$^{-3}$, $I_{2}$ is one simulated with the same treatment of
ionization and $v_{cut}$ = 10$^{-2}$. Data for this work are obtained
for sea level spectrum from Ref.~\protect\cite{bks1} 
(Eq.~(\protect\ref{bknsspec})); data from Ref.~\protect\cite{music1} 
represent result of simulation for spectrum from 
Ref.~\protect\cite{gaisser} (Eq.~(\protect\ref{gaisspec})).
}
\label{fig10n}
\end{figure}

Results reported in this Section are evidence of accuracy in parameterizations 
for muon cross sections and sea level spectrum to be the principal source of 
uncertainties when simulating atmospheric muon flux at depths where neutrino 
telescopes are located. It contributes uncertainty from 3\% 
(at the depth $D$ = 1 km in pure water) to 15\%
($D$ = 20 km) in resulting intensity of muon flux. Unfortunately, this
level has at present to be considered as a limit for accuracy of muon 
propagation algorithms. Influence of model for ionization exceeds this limit 
only for mono-energetic muon beams with initial energies $E \le$ 10 TeV and 
only if level of observation is at very last stage of muon range where major 
fraction of initial muon energy has been lost. Actually, due to steep shape of 
atmospheric muon power spectrum, an essential part of muons reaches detector 
location being  just on the last part of path. Therefore effect remains 
noticeable also for real atmospheric muons but in this case uncertainty was 
found to be much less: 2--3\%,
which is in an excellent agreement with Refs.~\cite{N94,music1}, while
Ref.~\cite{lagutin1} predicts much more significant difference (up to 17\%).
We suppose this disagreement may result from the fact that 
``small transfer grouping'' technique used in Ref.~\cite{lagutin1} treats muon 
cross sections to be constant between two interactions in contrast with the MUM
algorithm. In Ref.~\cite{music1} the same simplification was used but reported 
results were obtained by simulation with $v_{cut}$ = 10$^{-3}$. With such small 
$v_{cut}$ role of correct treatment for free path is not significant 
(see Sec.~\ref{sec:description} and Fig.~\ref{fig8n}). Choice of value for 
$v_{cut}$ is of even less importance and again, it is more critical if one 
investigates mono-energetic muon beam but for power spectrum alteration in 
$v_{cut}$ within $v_{cut} \le$ 0.05 leads only to 1--2\%
differences in simulated muon flux intensities. Up to $v_{cut}$ = 0.1
the error caused by rough account for fluctuations in energy losses remains
less than one which comes from uncertainties with muon energy loss. This 
conclusion is in a good agreement with level of errors reported in 
Ref.~\cite{music1}. Differences between muon flux intensities simulated for 
different models of ionization and values of $v_{cut}$, as obtained in given 
work and in Refs.~\cite{music1,lagutin1}, are presented in Fig.~\ref{fig9n} and 
Fig.~\ref{fig10n}.


\begin{figure}
\hspace{5.4cm}
\mbox{\epsfig{file=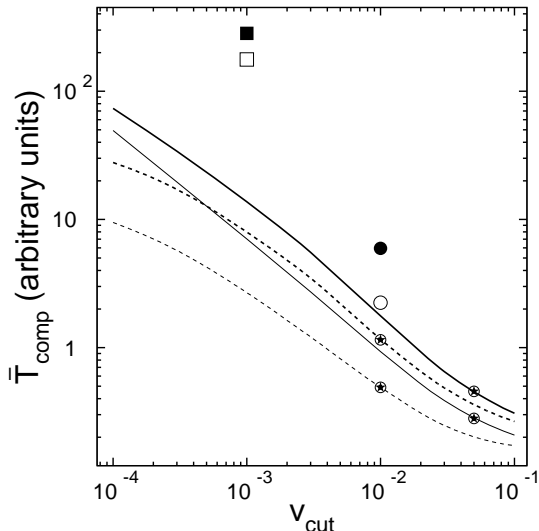,width=7.1cm}}
\protect\caption{
Averaged computation time $\bar T_{comp}$ which is necessary for muon 
propagation in pure water vs. $v_{cut}$, as obtained with the MUM code. Thick 
lines correspond to muon with initial energy $E_{s}$ = 9 TeV transported down 
to $D$ = 10 km. Thin lines are for $E_{s}$ = 1 TeV and $D$ = 3 km. Solid lines
show results for ionization included in SEL, dashed ones correspond to entirely
``continuous''  ionization. Circled asterisks on curves correspond to 
conservatively evaluated upper boundary for $v_{cut}$ below which the MUM 
algorithm inner accuracy has been proved to be high enough. This limit is equal
to $v_{cut}$ = 0.05 if ionization is included in SEL and to $v_{cut}$ = 0.01 if
ionization is entirely ``continuous'' (see Sec.~\protect\ref{sec:accuracy}). 
Circles and squares show values for $\bar T_{comp}$, as obtained with muon 
propagation codes PROPMU (version 2.01, 18/03/1993 with $v_{cut}$ = 10$^{-2}$ 
which is unchangeable) and MUSIC (version for pure water with bremsstrahlung
cross sections by Kehlner-Kokoulin-Petrukhin, 04/1999 with 
$v_{cut}$ = 10$^{-2}$ which is unchangeable), correspondingly. Closed markers 
are for $E_{s}$ = 9 TeV and $D$ = 10 km, open ones are for $E_{s}$ = 1 TeV and 
$D$ = 3 km.
}
\label{time}
\end{figure}

So when simulating muon fluxes at large depths with an ``ideal MC muon 
propagation algorithm'' it is reasonable to use $v_{cut} \approx$ 
0.05$\div$0.1 and entirely ``continuous'' model for ionization. Such setting of
simulation parameters does not lead to the error which would be out of 
insuperable uncertainties with muon energy loss but allows to save the 
computation time essentially. Fig.~\ref{time} show dependence of computation 
time on $v_{cut}$ and model for ionization, as was obtained with the MUM 
algorithm. Data for muon transportation codes PROPMU (Ref.~\cite{lipari}) and 
MUSIC (Ref.~\cite{music1}) are given on the figure, as well. We must emphasize 
that MUM in its presented version is 1D algorithm, in contrast both to PROPMU 
and MUSIC. PROPMU treats only Coulomb multiple scattering while in MUSIC the 
angle of the muon acquired in all radiative processes is also simulated which
takes an additional computation time. We evaluate the factor by which 
computation time with MUM would increase in case of extension up to 
3D-algorithm as $\sim$ 2. 

Accounting for data on real accuracy of current version of the MUM code (see 
Sec.~\ref{sec:accuracy}) which and data on computation time presented in 
Fig.~\ref{time} we conservatively consider $v_{cut} = 0.05$ and knock-on 
electron production included in SEL as an optimum setting for presented 
algorithm which allows to obtain the accurate results with relatively high 
speed. With such setting the proportion of computation time which is necessary 
to get the same statistics with MUM, PROPMU and MUSIC is approximately 
$1 : 10 : 600$. Of course, for some methodical purposes it may be necessary to 
choose more fine $v_{cut}$, e.g. if one wants to exclude an additional error 
when comparing results of simulations for different models of atmospheric muon 
sea level spectrum with each other or investigating survival probabilities 
which are much more sensitive to value of $v_{cut}$ than simulated spectrum
of atmospheric muons at large depths. 

We did not investigate specially the influence of simulation parameters on the 
results for the muon flux originated from neutrino but simple argumentation may
be applied for this case. In contrast with atmospheric muons whose source is 
far away of underwater, under-ice or underground detector and whose flux may 
only decrease when passing from the sea level down to detector depth, the 
source for muons which are produced in $\nu N$ interactions is uniformly 
distributed over water and/or rock both out- and inside the array. Intensity of
the muon flux $I^{ac}_{\mu}$ which accompanies the neutrino flux in a medium is 
proportional to the muon range, and, consequently, 
$I^{ac}_{\mu} \propto (dE/dx)_{total}^{-1}$ while simulated flux of atmospheric
muons at large depths depends more sharply upon muon energy loss as was shown 
in this Section. Thus, 
one may conclude that the setting of parameters described
above fits even better for propagation of muons originated from neutrino.

It is impossible to foresee all particular cases and give some strict
conformity between setting of parameters at muon MC propagation code
and problem to be solved. But we tried to present in this Section the whole
set of data which are necessary to chose the optimum set in each concrete
case.

\section{Selected results and comparison with other algorithms}
\label{sec:results1}

In this Section we present selected data on survival probabilities and 
atmospheric muon spectra deep underwater as simulated with MUM. To obtain 
atmospheric muon spectra we set $v_{cut}$ = 0.05. As was shown in 
Sec.~\ref{sec:accuracy} and Sec.~\ref{sec:simply} it does not distort results 
comparing to simulation with smaller values of $v_{cut}$. To compute survival
probabilities more delicate tuning was applied: $v_{cut}$ = 10$^{-3}$. In both
cases ionization was included in SEL. We compare our data with ones obtained 
with the PROPMU and MUSIC algorithms. Data simulated with PROPMU 
[version 2.01, 18/03/1993] ($v_{cut}$ = 10$^{-2}$) and PROPMU 
[version 2.1, 01/2000] (both with $v_{cut}$ = 10$^{-3}$ and 
$v_{cut}$ = 10$^{-2}$) are very close to each other, in all figures of this 
Section results from PROPMU [version 2.01, 18/03/1993] are presented. We used
[version for pure water with bremsstrahlung cross sections by 
Kelner-Kokoulin-Petrukhin, 04/1999] with $v_{cut}$ = 10$^{-3}$ for MUSIC. When 
comparing results on atmospheric muons at large depths obtained for pure and 
sea water the data are recalculated to each other using value $\rho$ = 1.027 
g cm$^-3$ as a sea water density (Ref.~\cite{higashi,DUMAND}). The difference 
between pure and sea water is negligible small for the muon propagation  if one
works in water equivalent units which was tested  by us up to slant depth 
$D$ = 10 km w.e. (see also Ref.~\cite{naumov1}).

\begin{figure}
\vspace{-6mm}
\hspace{3.2cm}
\mbox{\epsfig{file=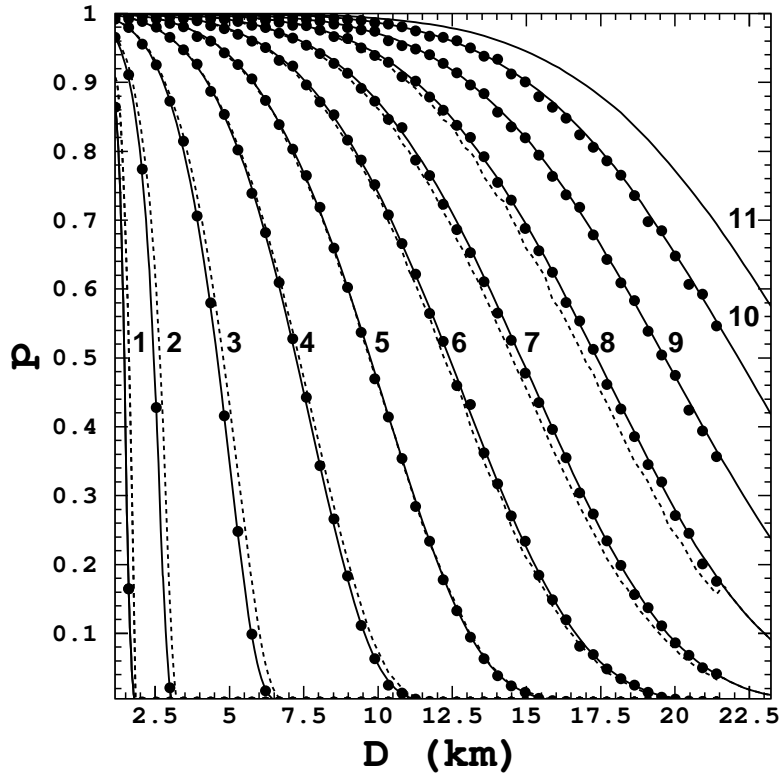,width=11.4cm}}
\protect\caption{
Survival probabilities vs. slant depth $D$ in pure water as computed with MUM
(solid lines), MUSIC (circles), and PROPMU (dashed lines). Figures near curves
indicate initial energies of muon beams which were as follows: 500 GeV (1),
1 TeV (2), 3 TeV (3), 10 TeV (4), 30 TeV (5), 100 TeV (6), 300 TeV (7),
1 PeV (8), 3 PeV (9), 10 PeV (10), 30 PeV (11). At simulations of data
presented on the plot muons are treated as stopped as soon as their energy 
decreases down to 10 GeV.
}
\label{klim}
\end{figure}

Fig.~\ref{klim} shows survival probabilities vs. slant depth $D$ in pure water 
as simulated for a set of initial muon beam energies from $E_{s}$ = 500 GeV to 
$E_{s}$ = 30 PeV. Survival probabilities obtained with MUM coincide within 
statistical errors with ones computed with MUSIC. PROPMU gives remarkably 
different values which are higher comparing to MUM and MUSIC output at muon 
energies $E_{s} \le$ 30 TeV and become less at $E_{s} >$ 30 TeV. 

\begin{figure}
\hspace{2.4cm}
\mbox{\epsfig{file=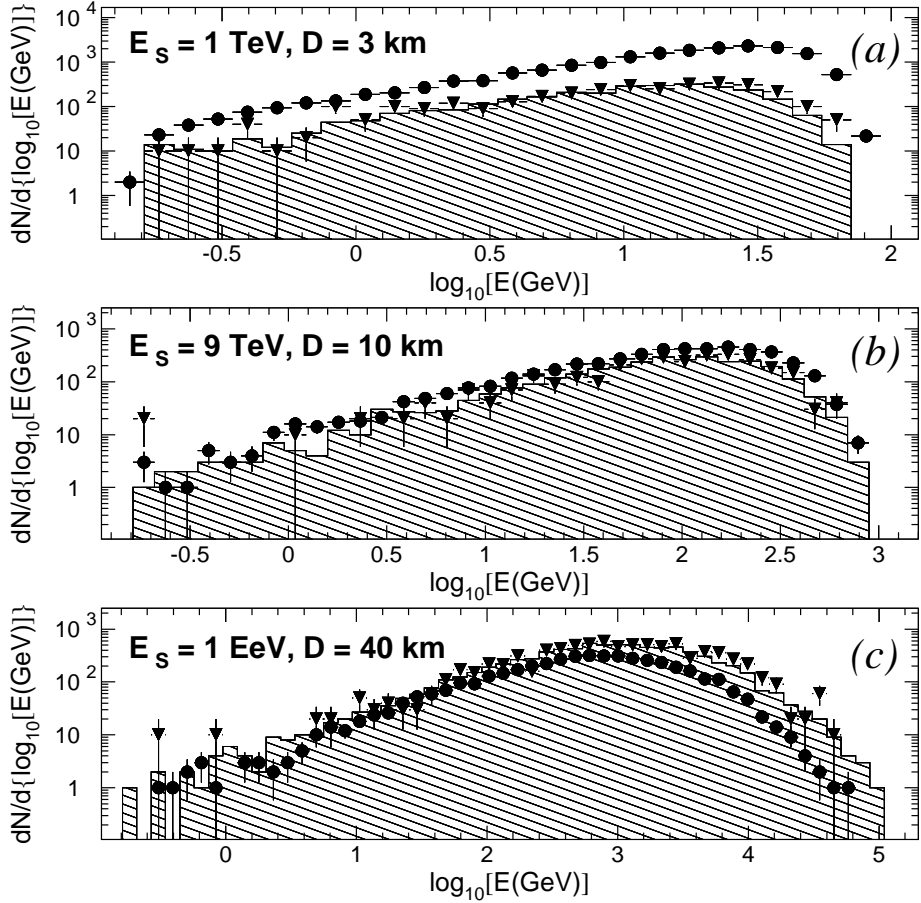,width=12.2cm}}
\protect\caption{Muon spectra resulting from mono-energetic muon beams with
initial energies $E_{s}$ = 1 TeV (a), $E_{s}$ = 9 TeV (b) and $E_{s}$ = 1 EeV
(c) after propagation down to depths $D$ = 3 km, 10 km and 40 km of pure 
water, correspondingly, as simulated with MUM (histograms), PROPMU (circles),
and MUSIC (triangles). Corresponding values for survival probabilities $p$ 
(fraction of muons survived after propagation) are equal to: 
$p$(1 TeV, 3 km) = 0.029 (MUM), 0.033 (MUSIC), 0.19 (PROPMU); 
$p$(9 TeV, 10 km) = 0.030 (MUM), 0.031 (MUSIC), 0.048 (PROPMU); 
$p$(1 PeV, 40 km) = 0.078 (MUM), 0.084 (MUSIC), 0.044 (PROPMU).
}
\label{sp}
\end{figure}

Fig.~\ref{sp} gives more detailed data on survival probabilities for three 
particular cases. It presents muon spectra resulted from mono-energetic muon 
beams with initial energies $E_{s}$ = 1 TeV (Fig.~\ref{sp}(a)), $E_{s}$ = 9 TeV 
(Fig.~\ref{sp}(b)) and $E_{s}$ = 1 PeV (Fig.~\ref{sp}(c)) after propagation
of distances 3 km, 10 km and 40 km in pure water, correspondingly. The 
distances were chosen so that survival probabilities would be much less than 1
in which case differences become more noticeable (see Sec.~\ref{sec:simply}).
A good agreement is observed between MUM and MUSIC data, while data 
obtained with PROPMU indicate essential differences which are of the same signs
as in Fig.~\ref{klim}.

\begin{figure}
\hspace{1.1cm}
\mbox{\epsfig{file=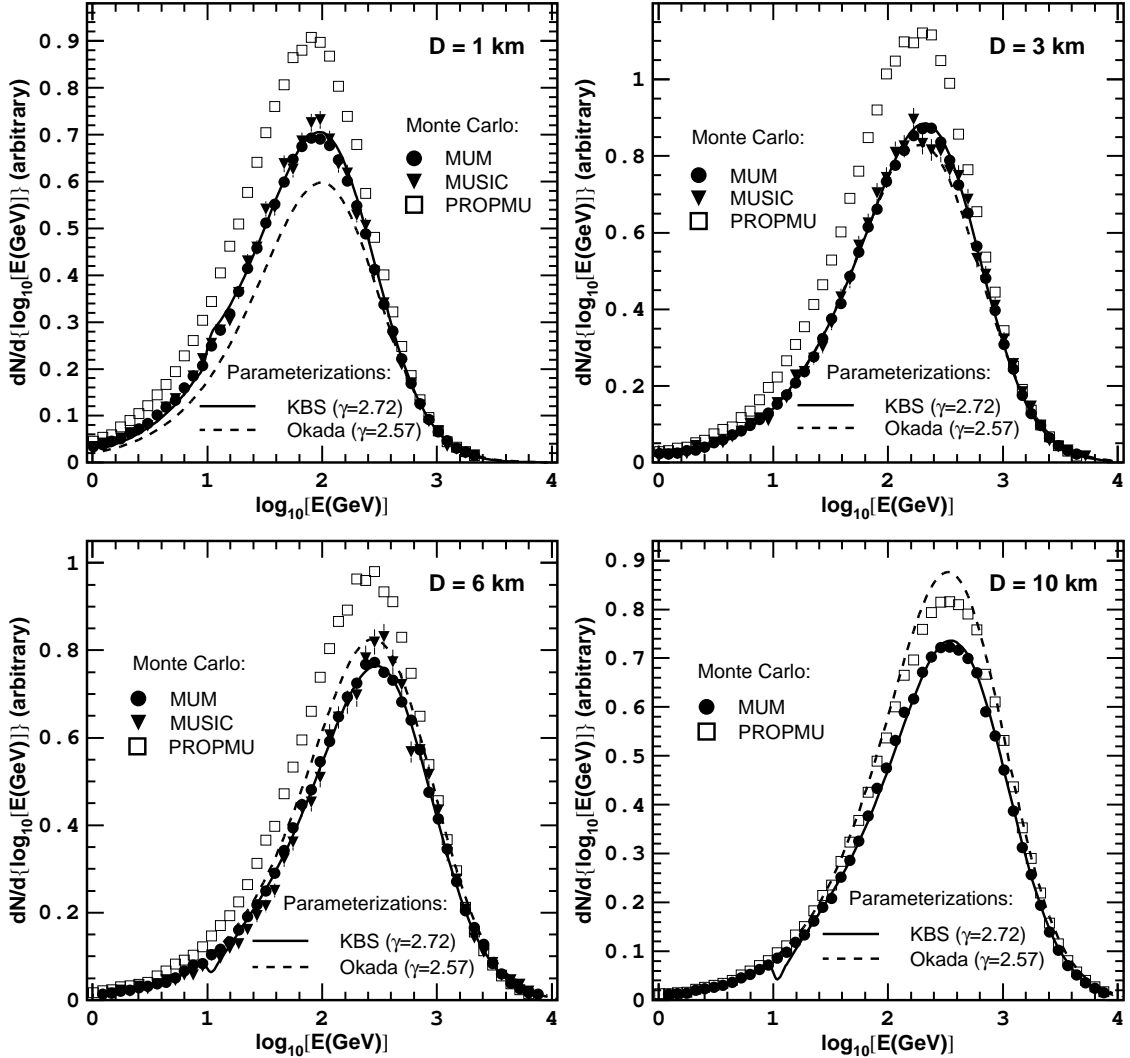,width=14.9cm}}
\protect\caption{
Differential spectra of vertical atmospheric muons at four depths in the pure
water as simulated with MUM, PROPMU and MUSIC (in all cases muon energies at
the sea level were sampled according to spectrum Eq.~(\ref{bknsspec})) and 
parameterized according to KBS with sea level spectrum Eq.~(\ref{bknsspec})
(Ref.~\protect\cite{bks1}) and A. Okada (Ref.~\protect\cite{okada}).
}
\label{sp1}
\end{figure}

In Fig.~\ref{sp1} differential spectra for vertical atmospheric muons at
different depths in pure water are presented as simulated with MUM, PROPMU and
MUSIC. Muons at the surface were sampled according to spectrum 
Eq.~(\ref{bknsspec}). 
Also parameterizations for deep underwater muon spectra as proposed by 
A.~Okada in Ref.~\cite{okada} and by S. Klimushin {\it et al.} in 
Ref.~\cite{bks1} (from here on will call them ``the Okada parameterization''
and ``the KBS parameterization'', correspondingly) are shown. The KBS
parameterization can adopt different models for sea level atmospheric muon 
spectrum. For data presented in Fig.~\ref{sp1} we used spectrum 
Eq.~(\ref{bknsspec}) which is basic one for the KBS parameterization. MUM gives 
almost the same results as MUSIC which could be expected because survival 
probabilities for muons in pure water are the same when simulating with MUSIC 
and MUM, as was shown above. Simulation with PROPMU produces the muon spectra 
which {\it (a)} are significantly higher (31\%, 30\%, 27\% and 17\%
in terms of integral muon flux at the depths $D =$ 1 km, 3 km, 6 km  and 10 km,
correspondingly) and {\it (b)} are expanded to the low energies. It is in good 
qualitative agreement with results on survival probabilities presented in 
Fig.~\ref{klim} and Fig.~\ref{sp}.
The coincidence between spectra simulated with MUM and curves for the basic KBS
parameterization results from the fact that in both cases the same sea level
atmospheric muon spectrum was adopted and, besides muon transport with the MUM
algorithm was applied to obtain the KBS parameterization. We would like to mark
that survival probabilities which KBS parameterization is based on 
were computed with
$v_{cut}$ = 10$^{-3}$. An excellent agreement with direct simulation in which 
$v_{cut}$ = 0.05 was set confirms the conclusion concerning insensitivity of
results on simulated atmospheric muon spectra at large depths on value of
$v_{cut}$ up to at least $v_{cut}$ = 0.05 (see Sec.~\ref{sec:simply}).
The Okada parameterization is lower than KBS, MUM and MUSIC results (up to 18\%
in terms of integral muon flux at $D =$ 1 km) at relatively shallow depths and
becomes higher at $D \ge$ 5 km because it is based on rather hard sea level
atmospheric muon spectrum with index $\gamma =$ 2.57 (Ref.~\cite{miyake}) which
leads to a deficit for low energy muons comparing to the 
basic KBS parameterization.

\begin{figure}
\hspace{3.8cm}
\mbox{\epsfig{file=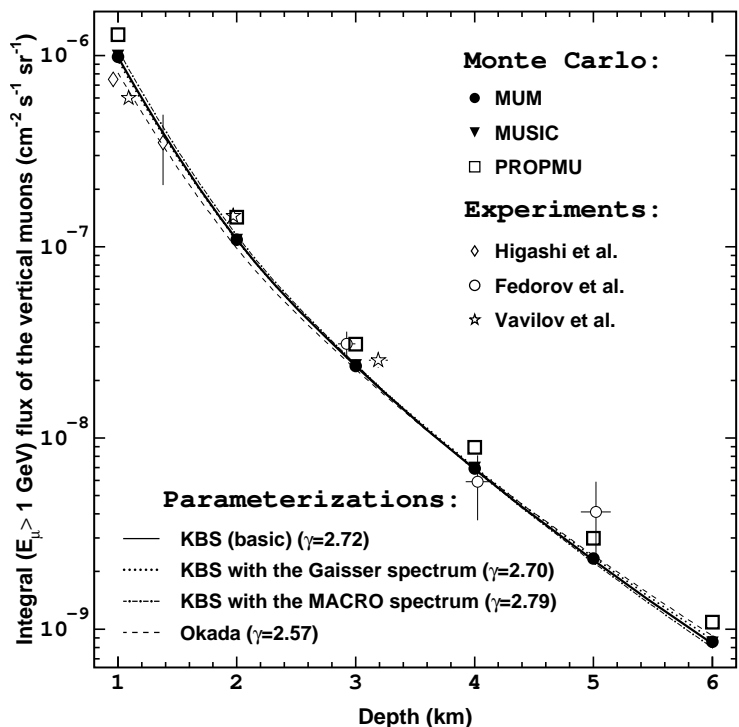,width=9.7cm}}
\protect\caption{
Results for integral flux of vertical atmospheric muons vs. depth in pure water
as {\it (1)} simulated with MUM, PROPMU and MUSIC with sea level spectrum
Eq.~(\protect\ref{bknsspec}); {\it (2)} parameterized by KBS 
(Ref.~\protect\cite{bks1}) with sea level atmospheric muon spectra 
Eq.~(\protect\ref{bknsspec}) (basic), from Ref.~\protect\cite{gaisser} (the 
Gaisser spectrum), from Ref.~\protect\cite{MACRO} (the MACRO spectrum) and 
A.~Okada (Ref.~\protect\cite{okada}) with sea level spectrum from 
Ref.~\protect\cite{miyake}; {\it (3)} measured by S.~Higashi
{\it et al.} (Ref.~\protect\cite{higashi}), V.~M.~Fedorov
{\it et al.} (Ref.~\protect\cite{fedorov}) and Yu.~N.~Vavilov
{\it et al.} (Ref.~\protect\cite{vavilov}).
}
\label{ig7}
\end{figure}

Fig.~\ref{ig7} presents results on integral flux of vertical atmospheric muons
at large depths in pure water as {\it (a)} simulated with MUM, PROPMU and MUSIC
for sea level spectrum Eq.~(\ref{bknsspec}); {\it (b)} parameterized by KBS 
(Ref.~\cite{bks1}) with sea level atmospheric muon spectra Eq.~(\ref{bknsspec})
(basic), from Ref.~\cite{gaisser} (the Gaisser spectrum), from Ref.~\cite{MACRO}
(the MACRO spectrum) and A.~Okada (Ref.~\cite{okada}) with sea level spectrum 
from Ref.~\cite{miyake}; {\it (c)} measured by S.~Higashi {\it et al.} 
(Ref.~\cite{higashi}), V.~M.~Fedorov {\it et al.} (Ref.~\cite{fedorov}) and 
Yu.~N.~Vavilov {\it et al.} (Ref.~\cite{vavilov}). Note that ``experimental'' 
points on the plot does not represent the pure experimental data because 
authors had to recalculate obtained counting 
rates to the vertical direction using
a model for the muon angular spectrum underwater. MUM and MUSIC results coincide
with each other within 1$\div$2\%.
Results from PROPMU algorithm exceed points from MUSIC and MUM by $\sim$ 30\%
being higher than any of presented parameterizations.

We also compared the data on muon propagation through the standard rock
obtained with MUM and MUSIC. Mean energy for vertically down-going
atmospheric muons sampled with sea level spectrum from Ref.~\cite{gaisser} was 
computed with MUM as $\bar E$ = 123$\pm$2 GeV, 256$\pm$4 GeV and 387$\pm$7 GeV
at depths $D$ = 1 km w.e., 3 km w.e. and 10 km w.e., respectively. 
The corresponding values simulated with the MUSIC code and
reported in Ref.~\cite{music1} are $\bar E$ = 125$\pm$1 GeV, 259$\pm$3 GeV 
and 364$\pm$4 GeV. So the maximum difference observed at the depth $D$ = 10 km 
w.e. is of 6\%.

Thus, 
results on survival probabilities and atmospheric muon spectra at large 
depths as simulated with MUM are practically in a coincidence with ones obtained
with MUSIC and are not also 
in contradiction with any experimental and theoretical
results presented in this section. The PROPMU algorithm shows noticeable
differences with MUM which are in good qualitative agreement to each other:
higher survival probabilities lead to higher muon fluxes deep underwater. It is
difficult to clarify the source of observed discrepancies without detailed
comparison for all steps of the algorithms but we believe that they can not be
explained only by a difference in models for muon energy loss as it is used 
in MUM (see Appendix \ref{app:cs}) and PROPMU (Refs.~\cite{lohman,lipari}) 
which does not exceed 2\% 
at $E \le$ 10 TeV (in terms of stopping power) being besides of both signs.

More data obtained with the MUM algorithm can be found in Ref.~\cite{bks1}.   

\section{Conclusions}
\label{sec:conclusions1}

We have presented the muon propagation Monte Carlo FORTRAN code MUM 
(MUons+Medium) and have given selected results obtained with the code for muon
spectra at large depths and survival probabilities in comparison with results 
obtained with other muon transportation algorithms. 
It was shown that for majority of applications it is quite enough to account
only for fluctuations in the radiative energy loss with fractions of energy 
lost as large as $v \ge v_{cut}$ = 0.05$\div$0.1 while ionization energy loss
may be entirely accounted by stopping power formula, as well as radiative 
energy loss with fractions of energy lost $v < v_{cut}$ = 0.05$\div$0.1. This
gives an essential advantage in terms of computation time comparing to commonly
used $v_{cut}$ = 10$^{-3}\div$10$^{-2}$ without lost of accuracy when 
simulating both propagation of atmospheric muons and muons which are born in
$\nu N$ interactions. However in practice it makes particular demands to 
accuracy of MC algorithm. Some customary simplifications (e.g. Eq~(\ref{Lap}))
which work perfectly when $v_{cut}$ = 10$^{-3}\div$10$^{-2}$ become sources
of significant errors when $v_{cut}$ increases. The sign and value of these
errors depend also on whether fluctuations in ionization are accounted or not.
So for presented version of the MUM algorithm the optimum set of simulation
parameters was conservatively evaluated by us (accounting results on inner
accuracy test and dependence of computation time on $v_{cut}$) to be
$v_{cut}$ = 0.05 and knock-electron production included in SEL.
                 
Our point of view on advantages of MUM is as follows. It is flexible enough
and provides with eventuality to tune parameters of simulation to an optimum 
for each concrete case to get desirable equilibrium between computation time 
and accuracy. Medium composition and parameterizations for muon cross sections 
are easily changeable. Inner accuracy of the code was conservatively evaluated 
to be 2$\times$10$^{-3}$ or better. Besides,  MUM provides with the special 
routine which allows to test inner accuracy for each given set of simulation 
parameters and take it into account when evaluating significance of results. 
The main disadvantage is that MUM in its reported version does not still treat
the three dimensions like, e.g. PROPMU and MUSIC codes. So it can not be used
to obtain lateral and angular deviations of muons at propagation through 
matter. Also other important features are still missed in MUM - for instance, 
treatment of composed medium as it is possible, e.g. in the latest version of 
the MUSIC code (Ref.~\cite{kud2}). But we consider the current version of MUM 
as a basis for further development and  plan to complement it, step by step, by
all necessary features.

The MUM code is available on request to {\it sokalski@pcbai10.inr.ruhep.ru}.

\acknowledgments

We would like to express our gratitude to I. Belolaptikov for useful discussions
which allowed to improve the MUM algorithm essentially. We are grateful to 
A.~Butkevich, R.~Kokoulin, V.~Kudryavzev, P.~Lipari, W.~Lohmann, V.~Naumov, 
O.~Streicher and Ch.~Wiebusch who read the paper at the draft study and gave 
their comments which mostly were taken into account both in the final version 
of the article and in the algorithm itself. We thank an anonymous
referee whose constructive and benevolent critique stimulated us for essential 
improvements in the article text. One of us (I.S.) was benefited a lot by 
attention and support of L.~Bezrukov and Ch.~Spiering. 

\appendix
\section{Parameterizations for muon cross sections used in the MUM algorithm}
\label{app:cs}

We use the following designations in this section:
$\alpha$ = 7.297353$\times$10$^{-3}$ -- fine structure constant;
$r_{e}$ = 2.817941$\times$10$^{-13}$ cm -- classical radii of  electron;
$m_{\mu}$ = 0.1056593 GeV and $m_{e}$ = 0.5110034 MeV -- muon and electron 
masses, correspondingly;
$N_{A}$ = 6.022$\times$10$^{23}$ -- the Avogadro number;
$Z$ and $A$ -- electric charge and atomic weight, correspondingly;
$e$ = 2.718282;
$\pi$ = 3.141593.
Other notations are explained in comments to formulas when necessary.

\subsection{Bremsstrahlung}
\label{app:brem}

We use differential cross section for bremsstrahlung as parameterized by
Yu. M. Andreev, L. B. Bezrukov and E. V. Bugaev in Ref.~\cite{brembb} as
a basic parameterization:

\[
\frac{d\sigma^{b}}{dv}(E,v) = \alpha\left(2r_eZ\frac{m_e}{m_\mu}\right)^2
\frac{1}{v}\left[\left(2-2v+v^2\right)\Psi_1\left(q_{\rm min},Z\right)
-\frac{2}{3}\left(1-v\right)\Psi_2\left(q_{\rm min},Z\right)\right],
\]
\[
\Psi_{1,2}\left(q_{\min},Z\right) =
                               \Psi_{1,2}^0\left(q_{\min},Z\right)
                              -\Delta_{1,2}\left(q_{\min},Z\right),
\]
\begin{eqnarray*}
\Psi_1^0\left(q_{\min},Z\right) & = & \frac{1}{2}\left(1+\ln
\frac{m_\mu^2a_1^2}{1+x_1^2}\right)-x_1\arctan\frac{1}{x_1}+
\frac{1}{Z}\left[\frac{1}{2}\left(1+\ln\frac{m_\mu^2 a_2^2}
                 {1+x_2^2}\right)-x_2\arctan\frac{1}{x_2}\right], \\
\Psi_2^0\left(q_{\min},Z\right) & = & \frac{1}{2}\left(\frac{2}{3}+
\ln\frac{m_\mu^2 a_1^2}{1+x_1^2}\right)+2x_1^2
\left(1-x_1\arctan\frac{1}{x_1}+\frac{3}{4}
                            \ln\frac{x_1^2}{1+x_1^2}\right)       \\
                     & + & \frac{1}{Z}\left[\frac{1}{2}
\left(\frac{2}{3}+\ln\frac{m_\mu^2 a_2^2}{1+x_2^2}\right)+
2x_2^2\left(1-x_2\arctan\frac{1}{x_2}+\frac{3}{4}\ln\frac{x_2^2}
                                    {1+x_2^2}\right)\right],
\end{eqnarray*}
\begin{eqnarray*}
\Delta_1\left(q_{\min},Z\ne1\right) & = & \ln\frac{m_\mu}{q_c}+
                       \frac{\zeta}{2}\ln\frac{\zeta+1}{\zeta-1},\\ 
\Delta_2\left(q_{\min},Z\ne1\right) & = & \ln\frac{m_\mu}{q_c}
            +\frac{\zeta}{4}\left(3-\zeta^2\right)\ln\frac{\zeta+1}
                             {\zeta-1}+\frac{2 m_\mu^2}{q_c^2},\\
\Delta_{1,2}\left(q_{\min},Z=1\right) & = & 0,
\end{eqnarray*}
\[
q_{\min} = \frac{m_\mu^2 v}{2E(1-v)}, \quad x_i = a_i q_{\min},
\]
\[
a_1   = \frac{111.7}{Z^{1/3}m_e}, \quad
a_2   = \frac{724.2}{Z^{2/3}m_e}, \quad
\zeta = \sqrt{1+\frac{4m_\mu^2}{q_c^2}}, \quad
q_c   = \frac{1.9 m_\mu}{Z^{1/3}}.
\]
Integration limits for bremsstrahlung in Eqs.  (\ref{CL}),
(\ref{L}), (\ref{sigmatot}) and (\ref{Li1}) are 
\[
v_{min}^{b} = 0, \quad
v_{max}^{b} = 1 - \frac{3}{4}\sqrt{e}\,(m_{\mu}/E)Z^{1/3}.
\]
Note that this parameterization does not account for contribution from
$e$-diagrams for bremsstrahlung when $\gamma$-quantum is emitted by atomic 
electrons which are knocked on by recoil (Ref.~\cite{bremkok}). 
Corresponding corrections are done in parameterizations for knock-on 
electron production (see Appendix \ref{app:elec}) according to 
Ref.~\cite{thankok}. Optionally, differential cross section for muon 
bremsstrahlung can be also treated in MUM according to parameterization 
given by S. R. Kelner, R. P. Kokoulin, and A. A. Petrukhin 
(Ref.~\cite{bremkok,thankok}).

\subsection{Photo-nuclear interaction}
\label{app:phnu}

We use parameterization for the photo-nuclear interaction of muon proposed by
L. B. Bezrukov and E. V. Bugaev (Ref.\cite{phnubb}):
\begin{eqnarray*}
\frac{d\sigma^{n}}{d v} &=& \frac{\alpha}{8\pi}A\sigma_{\gamma N}\,
                 v\left\{H(v)\ln\left(1+\frac{m_2^2}{t}\right)
                 -\frac{2m_\mu^2}{t}\left[1-\frac{0.25m_2^2}{t}
                 \ln\left(1+\frac{t}{m_2^2}\right)\right]\right.\\
                      &+& \left.G(z)\left[H(v)\left(\ln
                 \left(1+\frac{m_1^2}{t}\right)
                 -\frac{m_1^2}{m_1^2+t}\right)-\frac{2m_\mu^2}{t}
              \left(1-\frac{0.25m_1^2}{m_1^2+t}\right)\right]\right\},
\end{eqnarray*}
\[
H(v)=1-\frac{2}{v}+\frac{2}{v^2}, \qquad
\]
\begin{eqnarray*}
G(z) & = & \frac{9}{z}\left[\frac{1}{2}+\frac{(1+z)e^{-z}-1}{z^2}\right] \qquad (Z\ne1), \\
G(z) & = & 3 \qquad (Z=1),
\end{eqnarray*}
\[
z     = 0.00282A^{1/3}\sigma_{\gamma N},\quad
t     = \frac{m_\mu^2 v^2}{1-v}, \quad
m_1^2 = 0.54\;{\rm GeV}^2, \quad
m_2^2 = 1.80\;{\rm GeV}^2.
\]
Total cross section for absorption of a real photon of energy 
$\nu=s/2m_N=vE$ by a nucleon, $\sigma_{\gamma N}$, can be calculated in
MUM optionally according to either parameterization from 
Ref~\cite{phnubb} (basic):
\[
\sigma_{\gamma N}=[114.3+1.647 \ln^{2}(0.0213\,\nu)]\,\mu b,
\]
or by the ZEUS parameterization (Ref.~\cite{ZEUS}):
\[
\sigma_{\gamma N}=(63.5\,s^{0.097}+145\,s^{-0.5})\,\mu b,
\]
where $s$ and $\nu$ are expressed in GeV$^{2}$ and GeV, correspondingly.

Parameterization (Ref.~\cite{phnubb}) is valid for $\nu >$ 1 GeV so we use
values $v_{min}^{n} = 0.8/E$(GeV) and $v_{max}^{n}$ = 1 as integration 
limits in Eqs.  (\ref{CL}), (\ref{L}), (\ref{sigmatot}) and (\ref{Li1}). 
Note that results of integration were tested to be almost insensitive to
the lower limit in a wide range 
0.2/$E$(GeV) $\le v_{min}^{n} \le$ 1.5/$E$(GeV).

\subsection{Direct electron-positron pair production}
\label{app:pair}

Cross section for direct $e^{+}e^{-}$-pair production is used in MUM as 
parameterized by R. P. Kokoulin and A. A. Petrukhin 
(Refs.\cite{thankok,pairkok}):
\[
\frac{d\sigma^{p}}{dv}(E,v) = \alpha^{2}\frac{2}{3\pi}r_{e}^{2}Z\left(Z+\zeta(Z)\right)\frac{1-v}{v}\int\limits_{\rho}\left[\Phi_{e}+\left(m_{e}/m_{\mu}\right)^{2}\Phi_{\mu}\right]d\rho,
\]
\[
\Phi_{e}=\left\{\left[\left(2+\rho^{2}\right)\left(1+\beta\right)+\xi\left(3+\rho^{2}\right)\right]\ln\left(1+\frac{1}{\xi}\right)+\frac{1-\rho^{2}-\beta}{1+\xi}-\left(3+\rho^{2}\right)\right\}L_{e},
\]
\[
\Phi_{\mu}=\left\{\left[\left(1+\rho^{2}\right)\left(1+\frac{3}{2}\beta\right)-\frac{1}{\xi}\left(1-\rho^{2}\right)\left(1+2\beta\right)\right]\ln\left(1+\xi\right)+\frac{\xi\left(1-\rho^{2}-\beta\right)}{1+\xi}+\left(1-\rho^{2}\right)\left(1+2\beta\right)\right\}L_{\mu},
\]
\[
L_{e}=\ln\left[\frac{R\,Z^{-1/3}\sqrt{\left(1+\xi\right)\left(1+Y_{e}\right)}}{\displaystyle
1+\frac{2m_{e}\sqrt{e}R\,Z^{-1/3}\left(1+\xi\right)\left(1+Y_{e}\right)}{E\,v\left(1-\rho^{2}\right)}}\right]-\frac{1}{2}\ln\left[1+\left(\frac{3}{2}\frac{m_{e}}{m_{\mu}}Z^{1/3}\right)^{2}\left(1+\xi\right)\left(1+Y_{e}\right)\right],
\]
\[
L_{\mu}=\ln\left[\frac{\displaystyle
\frac{2}{3}\frac{m_{\mu}}{m_{e}}R\,Z^{-2/3}}{\displaystyle
1+\frac{2m_{e}\sqrt{e}R\,Z^{-1/3}\left(1+\xi\right)\left(1+Y_{e}\right)}{E\,v\left(1-\rho^{2}\right)}}\right],
\]
\[
Y_{e}=\frac{5-\rho^{2}+4\beta\left(1+\rho^{2}\right)}{2\left(1+3\beta\right)\ln\left(3+1/\xi\right)-\rho^{2}-2\beta\left(2-\rho^{2}\right)},
\]
\[
Y_{\mu}=\frac{4+\rho^{2}+3\beta\left(1+\rho^{2}\right)}{\displaystyle
\left(1+\rho^{2}\right)\left(\frac{3}{2}+2\beta\right)\ln\left(3+\xi\right)+1-\frac{3}{2}\rho^{2}},
\]
\[
\beta=\frac{v^{2}}{2\left(1-v\right)}, \quad 
\xi=\left(\frac{m_{\mu}v}{2m_{e}}\right)^{2}\frac{\left(1-\rho^{2}\right)}{\left(1-v\right)}.
\]
Here $\rho = (\epsilon^{+}-\epsilon^{-})/(\epsilon^{+}+\epsilon^{-})$ is 
the asymmetry coefficient of the energy distribution of $e^{+}e^{-}$-pair, 
$\epsilon^{+}$ and $\epsilon^{-}$ are positron and electron energies,
correspondingly. Limits for integration over $\rho$ are determined by:
\[
0 \le \, \mid \rho \mid \, \le \left(1-\frac{6m_{\mu}^{2}}{E^{2}\left(1-v\right)}\right)\sqrt{1-\frac{4m_{e}}{Ev}}.
\]
$R$ is a parameter determined by the value of radiation
logarithm ($R$ = 183 for Thomas-Fermi model and slightly depends upon
$Z$ for Hartrey-Fock model).
Its values are taken from 
Ref.~\cite{R}, where  $R$ has been calculated for different atoms according
to Hartrey-Fock model.
$\zeta(Z) \approx$ 1 takes into account the pair production in collisions
with electrons. Values from $\zeta(Z)$ are computed according to 
Refs.~\cite{thankok,zeta}. Integration limits for $j = p$ in Eqs. 
 (\ref{CL}), (\ref{L}), (\ref{sigmatot}) and (\ref{Li1}) are 
\[
v_{min}^{p} = \frac{4m_{e}}{E}, \quad
v_{max}^{p} = 1 - \frac{3}{4}\sqrt{e}\,(m_{\mu}/E)Z^{1/3}.
\]

\subsection{Knock-on electron production}
\label{app:elec}

Cross section for knock-on electron production is parameterized in the MUM
algorithm as follows:
\[
\frac{d\sigma^{e}}{dv}(E,v) = 2\pi r_{e}^{2} Z \frac{m_{e}}{E}\left(\frac{1}{v^{2}}-\frac{1}{v}\,\frac{E}{v_{max}^{e}}+\frac{1}{2}\right)\left(1+\Delta_{e\gamma}(E,v)\right),
\]
\[
v_{max}^{e}=\frac{2m_{e}E}{m_{\mu}^{2}+2m_{e}E}.
\]
$\Delta_{e\gamma}(E,v)$ represents the correction which takes into account 
$e$-diagrams for bremsstrahlung (Refs.~\cite{bremkok,thankok}) resulting in 
additional recoil electrons:
\[
\Delta_{e\gamma}(E,v)=\frac{\alpha}{2\pi}\ln\left(1+\frac{2vE}{m_{e}}\right)\left[\ln\left(\frac{4E^{2}\left(1-v\right)}{m_{\mu}^{2}}\right)-\ln\left(1+\frac{2vE}{m_{e}}\right)\right],
\]
value of $v_{max}^{e}$ is used also as upper integration limit in  Eqs.
(\ref{L}) and (\ref{sigmatot}) for $j = e$.

\subsection{Ionization}
\label{app:bebl}

Following Refs.~\cite{bremkok,thankok} we treat in the MUM code 
$e$-diagrams for bremsstrahlung as a part of ionization process. Therefore 
we have to use a bit modified formula for ionization:
\[
\left[\frac{dE}{dx}(E)\right]_{ion}=\frac{K}{\beta^2}\frac{Z}{A}\rho
\left[\ln\left(\frac{2 m_e p^2 E_{\max}}{m_\mu^2 \bar I^{2}}\right)
+\frac{E_{\max}^2}{4 E^2}-2\beta^2-\delta\right]+
\frac{N_{A}}{A_{eff}}\rho E \sum_{i=1}^{n}\! 
\left[
k_{i}\!\!\!\int\limits_{0}^{v_{max}^{e}}\!\!
\Delta_{e\gamma}(E,v)v\,dv
\right].
\]
Here $K$ = 0.1535 MeV\,g$^{-1}$\,cm$^{2}$, $p$ is the muon momentum,
$\beta = p/E$ is the muon velocity, $\rho$ is the material density, 
$\bar I$ is the mean ionization  potential,
\[
E_{\max} = (2m_e p^2)/(m_\mu^2+m_e^2+2m_e E)
\]
is the maximum energy transferable to an electron, $\delta$ is the 
density-effect correction which is treated according to Ref.~\cite{density}:
\[
\delta=\theta(X-X_0)\left[4.6052X+a\theta(X_1-X)(X_1-X)^m+C\right],
\]
where $\theta$ is the step function ($\theta(x) = 0$ at $x \leq 0$ and 
$\theta(x) = 1$ at $x > 0$), $X = \log_{10}(p/m_{\mu})$. The values $X_0$, 
$X_1$, $a$, $m$ and $C$ depend on the material and can be found in 
Refs.~\cite{lohman,density} along with values for $\bar I$, $\rho$ and 
$Z/A$. The first term represents Bethe-Bloch formula with corrections for 
density effect, the second one accounts bremsstrahlung $e$-diagrams. 
Expressions for $\Delta_{e\gamma}(E,v)$ and $v_{max}^{e}$ are given in
Appendix \ref{app:elec}, meaning of values $A_{eff}$ and $k_{i}$ is 
explained in Sec.~\ref{sec:description}.

\section{Free path between two muon interactions}
\label{app:path}

For the proof of the set of equation Eqs.~(\ref{one}) it is convenient
to introduce the kinetic equation for a propagation of a mono-energetic
muon beam through a medium. With the notations used in textbooks this
equation has the following view: 

\begin{equation}
\left\{
\begin{array}{rcl}
\partial n(E,t)/\partial t - \partial [\beta(E)n(E,t)] / \partial E
+ n(E,t) / \lambda(E) &=&0\\
n(E,0)&=&n_{0}\delta(E-E_{0})
\end{array}
\right.
\label{B1}
\end{equation}

\noindent
Here, $n(E,t)$ is the number of muons with energy $E$ after propagation
of distance $t$, $\beta(E)$ is the ``continuous'' energy loss per unit path,
$\lambda(E)$ is the muon mean free path  before interaction of SEL
type. The solution of Eq.~(\ref{B1}) is:

\begin{equation}
n(E,t)=n_{0}\delta(E-\epsilon(E_{0},t)) \exp \left[-\int\limits_{E}^{E_{0}}dE^{'}/\left(\lambda(E^{'})\beta(E^{'})\right)\right],
\label{B2}
\end{equation}

\noindent
where $\epsilon(E_{0},t)$ is found from the equation

\begin{equation}
\int\limits_{\epsilon(E_0,t)}^{E_0}dE/\beta(E) = t.
\label{B3}
\end{equation}

\noindent
Notice that Eq.~(\ref{B2}) can be rewritten as:

\begin{equation}
\eta(E,t)=\delta(E-\epsilon(E_{0},t)) \exp \left[-\int\limits_{E}^{E_{0}}dE^{'}/\left(\lambda(E^{'})\beta(E^{'})\right)\right],
\label{B4}
\end{equation}

\noindent
where $\eta(E,t)$ is the probability for a single muon to pass the path $t$
without interaction of SEL type and then one can easily see that 
Eqs. (\ref{B3}) and (\ref{B4}) lead to the Eqs.~(\ref{one})
after the following substitutions which are necessary for a return  to
the notations of Sec.~\ref{sec:description}:

\[
E_{0} \to E_{1}, \quad
E \to E_{2},\quad
\lambda(E) \to \bar L(E), \quad
\beta(E) \to \left[ dE(E)/dx\right]_{CEL}, \quad
t \to L.
\]

\newpage

\end{document}